\documentclass[acmsmall,screen]{acmart}

\setcopyright{rightsretained}
\acmPrice{}
\acmDOI{10.1145/3428295}
\acmYear{2020}
\copyrightyear{2020}
\acmSubmissionID{oopsla20main-p583-p}
\acmJournal{PACMPL}
\acmVolume{4}
\acmNumber{OOPSLA}
\acmArticle{227}
\acmMonth{11}

\usepackage{algorithm}
\usepackage[noend]{algpseudocode}
\usepackage{listings}
\usepackage{mathtools}
\usepackage{multirow}
\usepackage{tcolorbox}
\theoremstyle{definition}

\usepackage{color, soul}
\definecolor{grey}{rgb}{0.9,0.9,0.9}
\usepackage{balance}
\usepackage{textcomp}
\usepackage[T1]{fontenc}

\usepackage{listings} 
\usepackage{alltt}
\usepackage{microtype}
\usepackage{latexsym}
\usepackage{xspace}
\usepackage{url}
\usepackage{paralist}
\usepackage{pdfpages}
\usepackage{listings} 
\usepackage{alltt}
\usepackage{hyperref}
\usepackage{booktabs}  
\usepackage{subcaption} 
\usepackage{url}
\usepackage{hhline}
\usepackage{multicol}
\usepackage{makecell}
\usepackage{todonotes}
\usepackage[font={small}]{caption}
\usepackage{enumitem}
\usepackage{color}
\usepackage[export]{adjustbox}

\hypersetup{colorlinks,
  linkcolor=ACMDarkBlue,
  citecolor=ACMPurple,
  urlcolor=ACMDarkBlue,
  filecolor=ACMDarkBlue}

\xspaceaddexceptions{\}}

\newcommand{\suchthat}{{\mbox{s.t.\ }}}
\def\suchthatt{\,\middle|\,}

\newcommand{\set}[2]{\left\lbrace\,#1 \suchthatt #2\,\right\rbrace}

\newcommand{\scode}[1]{{\texttt{\small #1}}}

\newcommand{\scodetiny}[1]{{\texttt{\normalsize #1}}}
\newcommand{\tname}[1]{\textsc{#1}\xspace}

\newcommand{\euphony}{\tname{EuPhony}}

\newcommand{\bester}{\tname{Bester}}
\newcommand{\frangel}{\tname{FrAngel}}
\newcommand{\lasy}{\tname{LaSy}}
\newcommand{\concord}{\tname{Concord}}
\newcommand{\metal}{\tname{Metal}}
\newcommand{\dreamcoder}{\tname{DreamCoder}}
\newcommand{\ngds}{\tname{NGDS}}
\newcommand{\deepcoder}{\tname{DeepCoder}}
\newcommand{\cvc}{\tname{CVC4}}
\newcommand{\tool}{\tname{Probe}}
\newcommand{\stringbench}{\tname{String}}
\newcommand{\bitvec}{\tname{BitVec}}
\newcommand{\circuit}{\tname{Circuit}}
\newcommand{\sygus}{\tname{SyGuS}}
\newcommand{\eusolver}{\tname{EUSolver}}
\newcommand{\esolver}{\tname{ESolver}}

\newcommand{\ie}{\emph{i.e.\@}\xspace}

\newcommand{\eg}{\emph{e.g.\@}\xspace}

\newcommand{\vs}{\emph{vs.\@}\xspace}
\newcommand{\emphbf}[1]{\emph{\textbf{#1}\xspace}}
\newcommand{\mypara}[1]{\smallskip\noindent\emphbf{#1.}\xspace}

\newcommand{\examples}{\mathcal{E}}
\newcommand{\grammar}{\mathcal{G}}

\newcommand{\sem}[1]{
	\llbracket 
	#1 
	\rrbracket
}

\newcommand{\exec}{\ensuremath{\textsc{E}}}
\newcommand{\bank}{\ensuremath{\textsc{B}}}
\newcommand{\eval}{\ensuremath{\textsc{Eval}}}
\newcommand{\level}{\ensuremath{\textsc{Lvl}}}

\newcommand{\partialSols}{\ensuremath{\mathrm{PSol}}}
\newcommand{\costfn}{\icost}

\renewcommand{\gets}{\leftarrow}

\newcommand{\all}{\textsc{all }}
\newcommand{\prog}{P}
\newcommand{\round}[1]{\ensuremath{\lfloor#1\rceil}}
\newcommand{\nontermset}{\mathcal{N}}
\newcommand{\ruleset}{\mathcal{R}}
\newcommand{\start}{\mathcal{S}}
\newcommand{\termset}{\Sigma}
\newcommand{\rl}{\textsc{R}}
\newcommand{\nterm}{\textsc{N}}
\newcommand{\icost}{\mathrm{cost}}
\newcommand{\rcost}{\mathrm{rcost}}
\newcommand{\iarity}{\mathrm{arity}}
\newcommand{\fit}{\textsc{Fit}}
\newcommand{\pcfg}{p}
\newcommand{\llim}{\textsc{Lim}}

\algrenewcommand\algorithmicrequire{\textbf{Input:}}
\algrenewcommand\algorithmicensure{\textbf{Output:}}

\newif\ifdraft
\drafttrue

\ifdraft
\newcommand\authorrnote[3]{\textcolor{#1}{({#2}: {#3})}\xspace}
\newcommand{\TODO}[1]{{\color{orange!80!black}[\textsl{#1}]}}
\else
\newcommand\authorrnote[2]{}
\newcommand{\TODO}[1]{}
\fi

\newif\iflong
\longtrue

\xspace
\xspace

\newcommand{\many}[1]{\overrightarrow{#1}}

\newcommand{\nonterm}[1]{\textcolor{black}{#1}}

\newtheorem*{assumption*}{Assumption}

\newtheorem*{problem*}{Problem}

\newcommand{\ex}{$e_0$\xspace}
\newcommand{\exx}{$e_1$\xspace}
\newcommand{\exxx}{$e_2$\xspace}
\newcommand{\benchname}[1]{\scode{#1}\xspace}

\newcommand{\step}{\Rightarrow}
\newcommand{\manystep}{\Rightarrow^*}
\newcommand{\lang}{\mathcal{L}}
\newcommand{\vals}{\mathrm{Val}}
\newcommand{\trace}{\mathrm{tr}}

\begin{document}
	
	\title{Just-in-Time Learning for Bottom-Up Enumerative Synthesis}

	\author{Shraddha Barke}
	\affiliation{
		\institution{UC San Diego}
		\country{USA}  
	}
	\email{sbarke@eng.ucsd.edu}   
		
	\author{Hila Peleg}
	\affiliation{
		\institution{UC San Diego} 
		\country{USA}       
	}
	\email{hpeleg@eng.ucsd.edu}          
	
	\author{Nadia Polikarpova}
	\affiliation{
		\institution{UC San Diego}
		\country{USA}      
	}
	\email{npolikarpova@eng.ucsd.edu}   
	
  \begin{abstract}
  A key challenge in program synthesis is the astronomical size of the search space the synthesizer has to explore.
  In response to this challenge, recent work proposed to guide synthesis using learned probabilistic models.
  Obtaining such a model, however, might be infeasible for a problem domain where no high-quality training data is available.
  In this work we introduce an alternative approach to guided program synthesis:
  instead of training a model \emph{ahead of time}
  we show how to bootstrap one \emph{just in time}, during synthesis,
  by learning from partial solutions encountered along the way.
  To make the best use of the model,
  we also propose a new program enumeration algorithm we dub \emph{guided bottom-up search},  
  which extends the efficient bottom-up search with guidance from probabilistic models.
  
  We implement this approach in a tool called \tool,
  which targets problems in the popular syntax-guided synthesis (\sygus) format.
  We evaluate \tool on benchmarks from the literature
  and show that it achieves significant performance gains 
  both over unguided bottom-up search
  and over a state-of-the-art probability-guided synthesizer,
  which had been trained on a corpus of existing solutions.
  Moreover, we show that these performance gains do not come at the cost of solution quality:
  programs generated by \tool are only slightly more verbose than the shortest solutions
  and perform no unnecessary case-splitting. 
\end{abstract}

	\ccsdesc[500]{Software and its engineering~Domain specific languages}
	\ccsdesc[500]{Software and its engineering~Programming by example}
	
	\keywords{Program Synthesis, Probabilistic models, Domain-specific languages}  
	
	\maketitle
\section{Introduction}\label{sec:intro}

Consider the task of writing a program that satisfies examples in \autoref{fig:examples-intro}.
The desired program must return the substring of the input string \scode{s} 
on different sides of the dash, depending on the input integer \scode{n}. 
The goal of \emph{inductive program synthesis} is 
to perform this task automatically, \ie to generate programs from observations of their behavior.

\begin{figure}[h]
	\begin{tabular}{|cc|c|}
		\hline
		\multicolumn{2}{|c|}{\textbf{Input}}  & \textbf{Output} \\
		\scode{s} & \scode{n}  &   \\ \hline
		\scode{"1/17/16-1/18/17"} & \scode{1} & \scode{"1/17/16"}\\ \hline
		\scode{"1/17/16-1/18/17"} & \scode{2} & \scode{"1/18/17"}\\ \hline
		\scode{"01/17/2016-01/18/2017"} & \scode{1} & 
		\scode{"01/17/2016"} \\ \hline	
		\scode{"01/17/2016-01/18/2017"} & \scode{2} & 
		\scode{"01/18/2017"} \\ \hline		
	\end{tabular}
	\caption{Input-output example specification for the \benchname{pick-date} benchmark (adapted from \cite{euphony}).}\label{fig:examples-intro}
\end{figure}

Inductive synthesis techniques have made great strides in recent years~\cite{osera2015type,feser2015synthesizing,petrinetsynth17,FengMGDC17,shi2019frangel,gulwani2016pbe,wang2017synthesizing}, and are powering practical end-user programming tools~\cite{Gulwani:2011:ASP:1926385.1926423,PLDI-2014-LeG,InalaS18}.
These techniques adopt different approaches to perform search over the space of all programs 
from a \emph{domain-specific language} (DSL).
The central challenge of program synthesis is scaling the search to complex programs:
as the synthesizer considers longer programs, the search space grows astronomically large,
and synthesis quickly becomes intractable,  despite clever pruning strategies employed by state-of-the-art techniques. 

For example, consider the following solution to the \benchname{pick-date} problem introduced above,
using the DSL of a popular synthesis benchmarking platform \sygus~\cite{sygus}:
$$
\scode{(substr s (indexof (concat "-" s) "-" (- n 1)) (indexof s "-" n))}
$$
This solution extracts the correct substring of \scode{s}
by computing its starting index \scode{(indexof (concat "-" s) "-" (- n 1))}
to be either zero or the position after the dash, depending on the value of \scode{n}.
At size 14, this is the shortest \sygus program that satisfies the examples in \autoref{fig:examples-intro}.
Programs of this complexity already pose a challenge to state-of-the art synthesizers:
none of the \sygus synthesizers we tried were able to generate this or comparable solution within ten minutes%
\footnote{\cvc~\cite{reynolds2019cvc} is able to generate \emph{a} solution within a minute,
but its solution overfits to the examples and has size 73, which makes it hard to understand.}.

\mypara{Guiding synthesis with probabilistic models}
A promising approach to improving the scalability of synthesis
is to explore \emph{more likely programs first}.
Prior work~\cite{lee2018accelerating,balog2016deepcoder,menon2013machine,ellis2018search}
has proposed guiding the search using different types of learned probabilistic models.
%
For example, if a model can predict, given the input-output pairs in \autoref{fig:examples-intro}, 
that \scode{indexof} and \scode{substr} are more likely to appear in the solution than other string operations, 
then the synthesizer can focus its search effort on programs with these operations and find the solution much quicker.
Making this approach practical requires solving two major technical challenges:
\begin{inparaenum}[(1)]
\item \emph{how to obtain a useful probabilistic model}? and
\item \emph{how to guide the search} given a model?
\end{inparaenum}

\mypara{Learning a model}
Existing approaches~\cite{raychev2014code, bielik2016phog, lee2018accelerating} 
are able to learn probabilistic models of code automatically, 
but require significant amounts of high-quality training data, 
which must contain hundreds of meaningful programs \emph{per problem domain} targeted by the synthesizer.
Such datasets are generally difficult to obtain.

To address this challenge, we propose \emph{just-in-time learning},
a novel technique that learns a \emph{probabilistic context-free grammar} (PCFG)
for a given synthesis problem ``just in time'', \ie during synthesis, rather than ahead of time.
%
Previous work has observed~\cite{shi2019frangel,peleg2020perfect} that partial solutions---%
\ie programs that satisfy a subset of input-output examples---%
are often syntactically similar to the final solution.
Our technique leverages this observation to collect partial solutions it encounters during search 
and update the PCFG on the fly,
rewarding syntactic elements that occur in these programs.
For example, when exploring the search space for the \benchname{pick-date} problem,
unguided search quickly stumbles upon the short program \scode{(substr s 0 (indexof s "-" n)},
which is a partial solution, since it satisfies two of the four input-output pairs (with $\scode{n} = 1$).
At this point, just-in-time learning picks up on the fact that \scode{indexof} and \scode{substr}
seem to be promising operations to solve this problem, boosting their probability in the PCFG.  
Guided by the updated PCFG, our synthesizer finds the full solution in only 34 seconds.

\mypara{Guiding the search}
The state of the art in guided synthesis is \emph{weighted enumerative search} using the $A^*$ algorithm, 
implemented in the \euphony synthesizer~\cite{lee2018accelerating}
(see \autoref{sec:related} for an overview of other guided search techniques).
This algorithm builds upon \emph{top-down enumeration},
which works by gradually filling holes in incomplete programs.
Unfortunately, top-down enumeration is not a good fit for just-in-time learning:
in order to identify partial solutions, 
the synthesizer needs to \emph{evaluate} the programs it generates,
while with top-down enumeration the majority of synthesizer's time is spent generating incomplete programs that cannot (yet) be evaluated. 

To overcome this difficulty, we propose \emph{guided bottom-up search},
a new synthesis algorithm that, unlike prior work, 
builds upon \emph{bottom-up enumeration}.
This style of enumeration works by repeatedly combining small programs into larger programs;
every generated program is complete and can be evaluated on the input examples,
which enables just-in-time learning to rapidly collect a representative set of partial solutions.
In addition, bottom-up enumeration leverages dynamic programming 
and a powerful pruning technique known as \emph{observational equivalence}~\cite{udupa2013transit,albarghouthi2013recursive}, 
which further improves efficiency of synthesis.
Our algorithm extends bottom-up search with the ability to enumerate programs in the order of decreasing likelihood
according to a PCFG,
and to our knowledge, is the first guided version of bottom-up enumeration.
While guided bottom-up search enables just-in-time learning,
it can also be used with an independently obtained PCFG.

\begin{figure}[t]
	\includegraphics[scale=0.37]{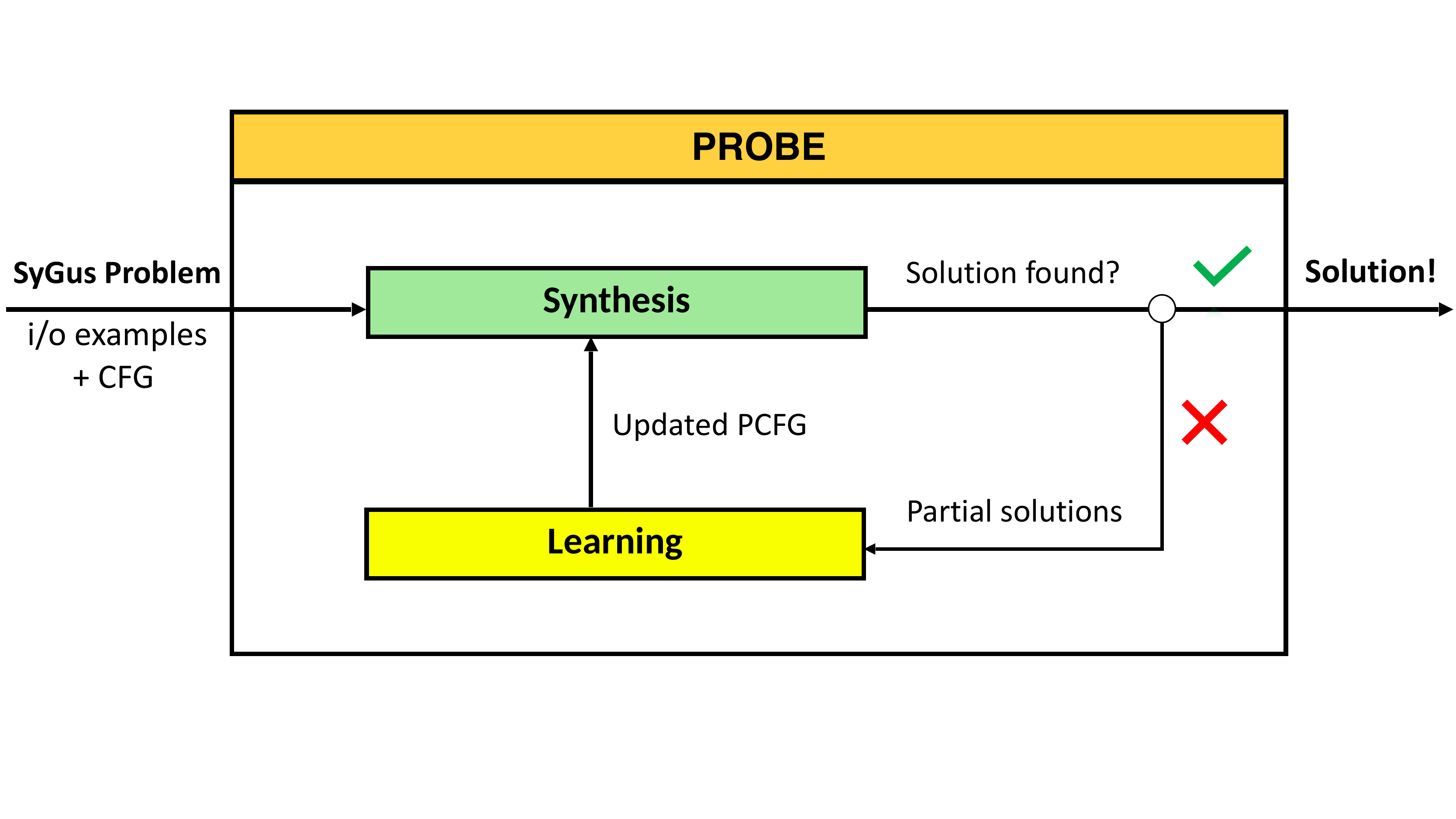}
	\vspace{-40pt}
	\caption{Overview of the \tool system}
	\label{fig:overview-probe}
\end{figure}

\mypara{The \tool tool}
We implemented guided bottom-up search with just-in-time learning in a synthesizer called \tool.
A high-level overview of \tool is shown in \autoref{fig:overview-probe}.
The tool takes as input an inductive synthesis problem in \sygus format,
\ie a context-free grammar of the DSL 
and a set of input-output examples%
\footnote{\tool also supports universally-quantified first-order specifications
and reduces them to input-output examples using counter-example guided inductive synthesis (CEGIS)~\cite{solar2006combinatorial}.
Since this reduction is entirely standard, in the rest of the paper we focus on inductive synthesis,
but we use both kinds of specifications in our evaluation.}; 
it outputs a program from the DSL that satisfies all the examples.
Optionally, \tool can also take as input initial PCFG probabilities suggested by a domain expert or learned ahead of time.
%

We have evaluated \tool on 140 \sygus benchmarks from three different domains: string manipulation, bit-vector manipulation, and circuit transformation. 
\tool is able to solve a total of 91 problems within a timeout of ten minutes, 
compared to only 44 problems for the baseline bottom-up synthesizer and 50 problems for \euphony.
Note that \tool outperforms \euphony despite requiring no additional training data,
which makes it applicable to new domains where large sets of existing problems are not available. 
We also compared \tool with \cvc~\cite{reynolds2019cvc}, the winner of the 2019 \sygus competition.
Although \cvc solves more benchmarks than \tool,
its solutions are less interpretable and tend to overfit to the examples:
\cvc solutions are 9 times larger than \tool solutions on average,
and moreover, on the few benchmarks where larger datasets are available,
\cvc achieves only 68\% accuracy on unseen data
(while \tool achieves perfect accuracy).

\mypara{Contributions}
To summarize, this paper makes the following contributions: 
\begin{enumerate}
	\item \emph{Guided bottom-up search:} 
  a bottom-up enumerative synthesis algorithm that explores programs in the order of decreasing likelihood defined by a PCFG
  (\autoref{sec:algo}).  
	\item \emph{Just-in-time learning:}
  a new technique for updating a PCFG during synthesis by learning from partial solutions
  (\autoref{sec:update}).
	\item \emph{\tool:}
  a prototype implementation of guided bottom-up search with just-in-time learning
  and its evaluation on benchmarks from prior work
  (\autoref{sec:eval}).
\end{enumerate}
\section{Background}\label{sec:prior}

\begin{figure}[t]
	\begin{tabular}{|l|l|l|}
		\hline
		\textbf{ID} & \textbf{Input}  & \textbf{Output}  \\ \hline
		\ex & \scode{"a < 4 and a > 0"} & \scode{"a  4 and a  0"}\\ \hline
		\exx & \scode{"<open and <close>"} & \scode{"open and close"}\\ \hline
		\exxx & \scode{"<Change> <string> to <a> number"} & 
		\scode{"Change string to a number"} \\ \hline		
	\end{tabular}
	\caption{Input-output example specification for the \benchname{remove-angles} benchmark (adapted from \cite{euphony}).}\label{fig:examples-running}
\end{figure}

In this section, we introduce the baseline synthesis technique that \tool builds upon: 
bottom-up enumeration with observational equivalence reduction~\cite{udupa2013transit,albarghouthi2013recursive}. 
For exposition purposes, hereafter we use a simpler running example than the one in the introduction;
the specification for this example, dubbed \benchname{remove-angles}, is given in \autoref{fig:examples-running}.
The task is to remove all occurrences of angle brackets from the input string. 

\subsection{Syntax Guided Synthesis}\label{sec:sygus}

We formulate our search problem as an instance of \emph{syntax-guided synthesis} (\sygus)~\cite{sygus}.
In this setting, synthesizers are expected to generate programs in a simple language of S-expressions 
with built-in operations on integers (such as $+$ or $-$) 
and strings (such as \scode{concat} and \scode{replace}). 
The input to a \sygus problem is a syntactic specification, 
in the form of a context-free grammar (CFG) that defines the space of possible programs 
and a semantic specification that consists of a set of input-output examples
\footnote{In general, \sygus supports a richer class of semantic specifications,
which can be reduced to example-based specifications using a standard technique, as we explain in \autoref{sec:eval}}. 
The goal of the synthesizer is to find a program generated by the grammar, 
whose behavior is consistent with the semantic specification.

\begin{figure}[t]
	\begin{math}\begin{array}{rrll}
	S &\rightarrow &\ \scode{arg}\ |\  \scode{"" } |\ \scode{"<" } |\ \scode{">"} & \quad\text{input string and string literals}\\
	& |&\ (\scode{replace}\ S\ S\ S) & \quad\text{\scode{replace s x y} replaces first occurrence of \scode{x} in \scode{s} with \scode{y}}\\
	& |&\ (\scode{concat}\ S\ S) & \quad\text{\scode{concat x y} concatenates \scode{x} and \scode{y}}\\
	\end{array}\end{math}
\caption{A simple CFG for string expressions and the informal semantics of its terminals.}\label{fig:ex-grammar}
\end{figure}%

For our running example \benchname{remove-angles},
we adopt a very simple grammar of string expressions shown in \autoref{fig:ex-grammar}.
The semantic specification for this problem is the set of examples $\{e_0, e_1, e_2\}$
from \autoref{fig:examples-running}. 
The program to be synthesized takes as input a string \scode{arg} 
and outputs this string with every occurrence of \scode{"<"} and \scode{">"} removed. 
Because the grammar in \autoref{fig:ex-grammar} allows no loops or recursion,
and the \scode{replace} operation only replaces the first occurrence of a given substring,
the solution involves repeatedly replacing the substrings \scode{"<"} and \scode{">"} 
with an empty string \scode{""}.
\autoref{tbl:replace-programs} shows one of the shortest solutions to this problem, 
which we dub \scode{replace-6}.
Note that this benchmark has multiple solutions of the same size that replace \scode{"<"} and \scode{">"} in different order;
for our purposes they are equivalent,
so hereafter we refer to any one of them as ``the shortest solution''.
The figure also shows two shorter programs, \scode{replace-2} and \scode{replace-3},
which satisfy different subsets of the semantic specification
and which we refer to throughout this and next section.


\begin{figure}[t]
	\resizebox{\textwidth}{!}{
	\begin{tabular}{|c|c|c|}
		\hline
		\textbf{ID}  & \textbf{Program} & \textbf{Examples Satisfied}\\ \hline
		\scode{replace-2} & \scodetiny{(replace (replace arg "<" "") ">" "")} & $\{e_0\}$  \\\hline
    \scode{replace-3} & \scodetiny{(replace (replace (replace arg "<" "") "<" "") ">" "")} & $\{e_0, e_1\}$ \\\hline
		\scode{replace-6} & \scodetiny{(replace (replace (replace (replace (replace (replace arg "<" "") "<" "") "<" "") ">" "") ">" "") ">" "")}& $\{e_0, e_1, e_2\}$\\  \hline
	\end{tabular}
	}
	\caption{Shortest solutions for different subsets of examples of the \benchname{remove-angles} problem.}\label{tbl:replace-programs}
\end{figure}

\subsection{Bottom-up Enumeration}\label{sec:height}

Bottom-up enumeration is a popular search technique in program synthesis,
first introduced in the tools \tname{Transit}~\cite{udupa2013transit} and \tname{Escher}~\cite{albarghouthi2013recursive}.
%
%
We illustrate this search technique in action 
using a simplified version of our running example, \benchname{remove-angles-short},
where the semantic specification only contains the examples $\{e_0, e_1\}$
(the shortest solution to this problem is the program \scode{replace-3} from \autoref{tbl:replace-programs}).


\mypara{Bottom-up Enumeration}
Bottom-up enumeration is a dynamic programming technique 
that maintains a \textit{bank} of enumerated programs 
and builds new programs by applying production rules to programs from the bank. 
\autoref{tbl:height} illustrates the evolution of the program bank on our running example.
Starting with an empty bank,
each iteration $n$ builds and adds to the bank all programs of height $n$.
In the initial iteration, we are limited to production rules that require no subexpressions---literals and variables; 
this yields the programs of height zero: \scode{"", "<", ">",} and \scode{arg}. 
In each following iteration,
we build all programs of height $n + 1$
using the programs of height up to $n$ as subexpressions.
For example at height one,
we construct all programs of the form $\scode{concat}\ x\ y$ and $\scode{replace}\ s\ x\ y$,
where $\langle s, x, y \rangle$ are filled with all combinations of height-zero expressions.
The efficiency of bottom-up enumeration
comes from reusing solutions to overlapping sub-problems, characteristic of dynamic programming:
when building a new program, all its sub-expressions are taken directly from the bank and never recomputed.


\begin{figure}[t]
	\resizebox{\textwidth}{!}{
	\begin{tabular}{|c|c|c|}
		\hline
		\textbf{Height}  & \textbf{$\#$ Programs} & \textbf{Bank} \\ \hline
		0 &  4 & \scode{arg, "", "<", ">"}\\  \hline
		\multirow{2}{*}{1} & \multirow{2}{*}{15} & \scodetiny{(concat arg arg), (concat arg "<"), (concat arg ">"), (concat "<" "<"), (concat "<" ">"), \ldots} \\
		& & \scodetiny{(replace arg "<" arg), (replace arg "<" ""), (replace arg "<" ">"), (replace arg ">" "<"), \ldots} \\ \hline
		\multirow{2}{*}{2} & \multirow{2}{*}{1023} & \scodetiny{(concat arg (concat arg arg)), (concat arg (concat ">" ">")), \ldots (concat "<" (concat arg arg)), } \\
		& & \scodetiny{(concat "<" (replace arg "<" arg)), (concat ">" (concat "<" "<")), (concat ">" (replace arg ">" "<"))} \\ \hline
		\multirow{2}{*}{3} & \multirow{2}{*}{ $\sim30M$} & \scodetiny{(concat arg (concat (replace arg "<" arg) arg)), (concat arg (concat (replace arg "<" arg) "<"))} \\
		& & \scodetiny{(concat arg (concat (replace arg "<" arg) ">")), (concat arg (concat (replace arg "<" "") arg)) $\ldots$} \\ \hline
	\end{tabular}
	}
	\caption{Programs generated for \benchname{remove-angles-short} from the grammar in \autoref{fig:ex-grammar} in the order of height.}\label{tbl:height}
\end{figure}

\mypara{Observational Equivalence Reduction}
Bottom-up synthesizers further optimize the search 
by discarding programs that are \emph{observationally equivalent} to some program that is already in the bank.
Two programs are considered observationally equivalent if they evaluate to the same output 
for every input in the semantic specification.
In our example, the height-one program \scode{(concat arg "")} is not added to the bank 
because it is equivalent to the height-zero program \scode{arg}. 
This optimization shrinks the size of the bank at height one from 80 to 15;
because each following iteration uses all combinations of programs from the bank as subexpressions,
even a small reduction in bank size at lower heights leads to a significant overall speed-up.

Despite this optimization, the size of the bank grows extremely quickly with height, as illustrated in \autoref{tbl:height}. 
In order to get to the desired program \scode{replace-3}, which has height three, 
we need to enumerate anywhere between 1024 and  $\sim30M$ programs 
(depending on the order in which productions and subexpressions are explored within a single iteration).
Because of this search space explosion, 
bottom-up enumerative approach does not find \scode{replace-3} even after 20 minutes.

\section{Our approach}\label{sec:overview}

In this section, we first modify bottom-up search
to enumerate programs in the order of \emph{increasing size} rather than \emph{height} (\autoref{sec:size})
and then generalize it to the order of \emph{decreasing likelihood} defined by a probabilistic context-free grammar (\autoref{sec:weighted-bottom-up}).
Finally, we illustrate how the probabilistic grammar can be learned \emph{just in time}
by observing partial solutions during search (\autoref{sec:just-in-time}).

\subsection{Size-based Bottom-up Enumeration}\label{sec:size}

Although exploring \emph{smaller programs first} is common sense in program synthesis,
the exact interpretation of ``smaller'' differs from one approach to another.
%
As we discussed in \autoref{sec:prior},
existing bottom-up synthesizers explore programs in the order of increasing height;
at the same time, synthesizers based on other search strategies~\cite{sygus,KoukoutosKK16,alur2017scaling}
tend to explore programs in the order of increasing \emph{size}---\ie total number of AST nodes---rather than height,
which has been observed empirically to be more efficient.

To illustrate the difference between the two orders, consider a hypothetical size-based bottom-up synthesizer.
\autoref{tbl:size} shows how the bank would grow with each iteration on our running example.
The solution \scode{replace-3} that we are looking for has size ten (and height three).
Hence, size-based enumeration only has to explore up to 2048 programs to discover this solution
(compared with up to  $\sim30M$ for height-based enumeration).
This is not surprising: a simple calculation shows that programs of height three range in size from 8 to 26,
and our solution is towards the lower end of this range;
in other words, \scode{replace-3} is tall and skinny rather than short and bushy.
This is not a mere coincidence:
in fact, prior work~\cite{shahscalable} has observed that \emph{useful programs tend to be skinny rather than bushy},
and therefore exploration in the order of size has a better inductive bias.


\begin{figure}[t]
	\resizebox{\textwidth}{!}{
		\begin{tabular}{|c|c|c|}
			\hline
			\textbf{Size}  & \textbf{$\#$ Programs} & \textbf{Bank} \\ \hline
			1 & 4 & \scode{arg, "", "<", ">"}\\  \hline
			2 & 0 & \scode{None}\\  \hline
			
			\multirow{2}{*}{3} & \multirow{2}{*}{9} & {\scode{(concat arg arg), (concat arg "<"), (concat arg ">"), (concat "<" arg),}} \\
			& & \scode{(concat "<" "<"), (concat "<" ">"), (concat ">" arg), (concat ">" "<"), (concat ">" ">")} \\ \hline
	
			\multirow{2}{*}{4} & \multirow{2}{*}{6} & {\scodetiny{(replace arg "<" arg), (replace arg "<" "), 
				(replace arg "<" ">")}}\\
			& & {\scodetiny{(replace arg ">" arg), (replace arg ">" ""),
			(replace arg ">" "<")}} \\ \hline
			
			\vdots & \vdots & \vdots\\  \hline
			\multirow{2}{*}{8} & \multirow{2}{*}{349} & {\scodetiny{ (concat (concat (replace arg "<" arg) arg) arg), (concat (replace arg ">" (concat ">" arg)) ">"), }}\\
			& & {\scodetiny{(replace (concat arg "<") (concat ">" "<") "") $\ldots$ (replace (concat ">" arg) (concat ">" "<") ">")}} \\ \hline
			
			\multirow{2}{*}{9} & \multirow{2}{*}{714} & \scodetiny{(concat (concat arg arg) (concat (concat arg arg) arg)), (concat (concat "<" "<") (concat (concat ">" "<") "<")), $\ldots$} \\
			& & {\scodetiny{ (replace (replace arg "<" "") "<" (concat ">" ">")),  (replace (replace arg ">" (concat ">" ">")) "<" ">")}} \\ \hline
			
			\multirow{2}{*}{10} & \multirow{2}{*}{2048} & \scode{(concat "<" (replace (concat arg arg) (concat ">" arg) "<")),} \\
			& & {$\ldots$ \scode{(concat arg (concat (replace arg "<" (concat ">" ">")) ">"))}} \\ \hline
		\end{tabular}
	}  
	\caption{Programs generated for \benchname{remove-angles-short} from the grammar in \autoref{fig:ex-grammar} in the order of size.}\label{tbl:size}
\end{figure}

\mypara{Extending bottom-up enumeration}
Motivated by this observation, we extend the bottom-up enumerative algorithm from \autoref{sec:height} 
to explore programs in the order of increasing size.
To this end, we modify the way subexpressions are selected from the bank in each search iteration.
For example, to construct programs of size four of the form $\scode{concat}\ x\ y$,
we only replace $\langle x, y \rangle$ with pairs of programs whose sizes add up to three
(the \scode{concat} operation itself takes up one AST node).
This modest change to the search algorithm yields surprising efficiency improvements:
our size-based bottom-up synthesizer is able to solve the \benchname{remove-angles-short} benchmark in only one second!
(Recall that the baseline height-based synthesizer times out after 20 minutes).

Unfortunately, the number of programs in the bank still grows exponentially with program size,
limiting the range of sizes that can be explored efficiently:
for example, the solution to  the original \benchname{remove-angles} benchmark (\scode{replace-6})
has size 19,
and size-based enumeration is unable to find it within the 20 minute timeout.
This is where \emph{guided bottom-up search} comes to the rescue.
\subsection{Guided Bottom-up Search}\label{sec:weighted-bottom-up}

Previous work has demonstrated significant performance gains in synthesizing programs 
by exploiting probabilistic models to guide the search~\cite{balog2016deepcoder,lee2018accelerating,menon2013machine}. 
These techniques, however, do not build upon bottom-up enumeration,
and hence cannot leverage its two main benefits: 
reuse of subprograms and observational equivalence reduction (\autoref{sec:height}).
Our \emph{first key contribution} is modifying the size-based bottom-up enumeration technique from previous section 
to guide the search using a \emph{probabilistic context-free grammar} (PCFG).
We refer to this modification of the bottom-up algorithm  as \textit{guided bottom-up search}. 

\mypara{Probabilistic context-free grammars}
A PCFG assigns a probability to each production rule in a context-free grammar.
For example, \autoref{fig:prob-grammar} depicts a PCFG for our running example
that is biased towards the correct solution:
it assigns high probabilities to the rules (operations) that appear in \scode{replace-6}
and a low probability to the rule \scode{concat} that does not appear in this program.
As a result, this PCFG assigns a higher likelihood to the program \scode{replace-6}%
\footnote{The likelihood of a program is the product of the probabilities of all rules involved in its derivation.}
than it does to other programs of the same size.
Hence, an algorithm that explores programs in the order of decreasing likelihood
would encounter \scode{replace-6} sooner than size-based enumeration would.
%


\begin{figure}[t]
	\begin{math}\begin{array}{rrllll}
  & & & p_R & -\log(p_R) & \icost_R\\  
	S &\rightarrow &\ \scode{arg}\ |\  \scode{""}\ |\ \scode{"<"}\ |\ \scode{">"} &  0.188 & 2.41 & 2 \\
	& |&\ (\scode{replace}\ S\ S\ S) & 0.188 & 2.41 & 2\\
	& |&\ (\scode{concat}\ S\ S) & 0.059 & 4.09 & 4\\
	\end{array}\end{math}
	\caption{A PCFG for string expressions that is biased towards the solution \scode{replace-6}.
  For each production rule $R$, we show its probability $p_R$ and its cost $\icost_R$,
  which is computed as a rounded negative log of the probability.}\label{fig:prob-grammar}
\end{figure}


\mypara{From probabilities to discrete costs}
Unfortunately, size-based bottom-up enumeration cannot be easily adapted to work with real-valued probabilities.
We observe, however, that the order of program enumeration need not be exact:
enumerating \emph{approximately} in the order of decreasing likelihood still benefits the search.
Our insight therefore is to convert rule probabilities into discrete \emph{costs},
which are computed as their rounded negative logs. 
According to \autoref{fig:prob-grammar},
the high-probability rules have a low cost of two, and the low-probability rule \scode{concat} has a higher cost of four.
The cost of a program is computed by summing up the costs of its productions, for example:
\begin{align*}
\icost(\scode{concat arg "<"}) &= \icost(\scode{concat}) + \icost(\scode{arg}) + \icost(\scode{"<"})\\ 
&=  4 + 2 + 2 = 8
\end{align*}
Hence, the order of increasing cost approximately matches the order of decreasing likelihood.

\mypara{Extending size-based enumeration}
With the discrete costs at hand,
guided bottom-up search is essentially the same as the size-based search detailed in \autoref{sec:size}, 
except that it takes the cost of the top-level production into account when constructing a new program. 
\autoref{tbl:size-vs-prob} illustrates the working of this algorithm.
For example, at cost level $8$, we build all programs of the form $\scode{concat}\ x\ y$,
where the costs of $x$ and $y$ sum up to $8 - 4 = 4$.
The cost of our solution \scode{replace-6} is 38,
which places it within the first 130K programs the search encounters;
on the other hand, its size is 19,
placing it within the first $\sim4M$ programs in the order of size.
As a consequence, size-based enumeration cannot find this program within 20 minutes,
but guided enumeration, given the PCFG from \autoref{fig:prob-grammar},
is able to discover \scode{replace-6} within 5 seconds.

\begin{figure}[t]
	\resizebox{\textwidth}{!}{

\begin{tabular}{|c|c|c|}
	\hline
	\textbf{Cost}  & \textbf{$\#$ Programs} & \textbf{Bank} \\ \hline
	2 &  4 & \scode{arg, "", "<", ">"}\\  \hline
	\multirow{2}{*}{8} & \multirow{2}{*}{15} & \scodetiny{(replace arg "<" arg), (replace arg "<" ""), (replace arg "<" ">")} \\
	& & \scodetiny{(replace arg ">" "<"), (concat "<" arg), (concat "<" "<") \ldots} \\ \hline
	\multirow{2}{*}{20} & \multirow{2}{*}{1272} & \scodetiny{(replace "<" (replace arg (replace arg "<" "") "") ""), (replace "<" (replace arg (replace arg "<" "") "") ">") \ldots} \\
	& & \scodetiny{(replace (replace arg ">" "<") (replace arg ">" "") arg), (replace (replace arg ">" "<") (replace arg ">" "") ">")} \\ \hline
			\vdots & \vdots & \vdots \\ \hline
	\multirow{2}{*}{38} & \multirow{2}{*}{130K} & \scodetiny{(str.replace (replace arg "<" (replace (replace arg ">" "<") ">" arg)) (replace (replace arg "<" "") ">" arg) "<")} \\
	& & \scodetiny{(replace (replace arg "<" (replace (replace arg ">" "<") ">" arg)) (replace (replace arg "<" "") ">" arg) ">") $\ldots$} \\ \hline
\end{tabular}
	}
	\caption{Programs generated for \benchname{remove-angles} using guided bottom-up search with the PCFG in \autoref{fig:prob-grammar}}\label{tbl:size-vs-prob}
\end{figure}


\subsection{Just-in-Time Learning}\label{sec:just-in-time}

In the previous section we have seen that guided bottom-up search can find solutions efficiently,
given an appropriately biased PCFG.
But how can we obtain such a PCFG for each synthesis problem?
Prior approaches have proposed learning probabilistic models
from a corpus of existing solutions~\cite{menon2013machine,lee2018accelerating}
(see \autoref{sec:related} for a detailed discussion).
%
While achieving impressive results, 
these approaches are computationally expensive
and, more importantly, require high-quality training data, 
which is generally hard to obtain.
Can we benefit from guided search when training data is not available?

Our \emph{second key contribution} is a new approach to learning probabilistic models of programs,
which we dub \textit{just-in-time learning}.
This approach is inspired by an observation made in prior work~\cite{shi2019frangel,peleg2020perfect}
that \emph{partial solutions}---programs that satisfy a subset of the semantic specification---%
often share syntactic similarity with the full solution.
We can leverage this insight
to iteratively bias the PCFG during synthesis,
rewarding productions that occur in partial solutions we encounter.


\mypara{Enumeration with just-in-time learning}
We illustrate just-in-time learning on our running example \benchname{remove-angles}.
We begin enumeration with a uniform PCFG,
which assigns the same probability to each production%
\footnote{The algorithm can also be initialized with a pre-learned PCFG if one is available.}.
In this initial PCFG every production has cost 3 (see \autoref{tbl:readjust-weight}).  

With a uniform PCFG, our search starts off exactly the same as size-based search of \autoref{sec:size}.
At size 7 (cost level 21), the search encounters the program \scode{replace-2},
which satisfies the example $e_0$.
Since this program contains productions \scode{replace}, \scode{arg}, \scode{""}, \scode{">"}, and \scode{"<"},
we \emph{reward} these productions by decreasing their cost,
as indicated in \autoref{tbl:readjust-weight};
after this update, the cost of the production \scode{concat} does not change, so our solution is now cheaper relative to other programs of the same size.
With the new PCFG at hand, the enumeration soon encounters another partial solution,
\scode{replace-3}, which covers the examples $e_0$ and $e_1$.
Since this program uses the same productions as \scode{replace-2} and satisfies even more examples,
the difference in cost between the irrelevant production \scode{concat} and the relevant ones increases even more:
in fact, we have arrived at the same biased PCFG we used in \autoref{sec:weighted-bottom-up} to illustrate the guided search algorithm.
%


\begin{figure}[t]
	\begin{tabular}{|c|c|c|}
		\hline
		\textbf{Partial Solution}  & \textbf{Examples Satisfied}  & \textbf{PCFG costs}\\ \hline
		 & $\emptyset$ & $\scode{arg}, \scode{""}, \scode{"<"}, \scode{">"}, \scode{replace}, \scode{concat}  \mapsto 3$ \\\hline
		\scode{replace-2} & $\{e_0\}$ & $\scode{arg}, \scode{""}, \scode{"<"}, \scode{">"}, \scode{replace}  \mapsto 2; \scode{concat} \mapsto 3$  \\\hline
    \scode{replace-3} & $\{e_0, e_1\}$ & $\scode{arg}, \scode{""}, \scode{"<"}, \scode{">"}, \scode{replace}  \mapsto 2; \scode{concat} \mapsto 4$ \\\hline
	\end{tabular}
	\caption{Just-in-time learning: as the search encounters partial solutions that satisfy new subsets of examples,
  PCFG costs are adjusted and the relative cost of \scode{concat}, which is not present in the solution, increases.}\label{tbl:readjust-weight}
\end{figure}


\mypara{Challenge: selecting promising partial solutions}
As this example illustrates, 
the more partial solutions we encounter that are similar to the final solution,
the more biased the PCFG becomes, gradually steering the search in the right direction.
The key challenge with this approach is 
that the search might encounter hundreds or thousands of partial solutions, 
and many of them have irrelevant syntactic features. 
%
In our running example, there are in fact more than 3100 programs that satisfy at least one of the examples $e_0$ or $e_1$. 
For instance, the program
\[\scode{replace (replace (replace (concat arg "<") "<" "") "<" "") ">" ""}\] 
satisfies $e_0$, but contains the \scode{concat} production,
so if we use this program to update the PCFG,
we would steer the search away from the final solution.
Hence, the core challenge is to identify \emph{promising} partial solutions,
and only use those to update the PCFG.

A closer look at this program reveals
that it has the same behavior as the shorter program \scode{replace-2},
but it contains an irrelevant subexpression that appends \scode{"<"} to \scode{arg}
only to immediately replace it with an empty string!
In our experience, this is a common pattern:
whenever a partial solution $p'$ is \emph{larger} than another partial solution $p$
but solves the same subset of examples,
then $p'$ often syntactically differs from $p$ by an irrelevant subexpression,
which happens to have no effect on the inputs solved by the two programs.
Following this observation, we only consider a partial solution $p$ promising---%
and use it to update the PCFG---%
when it is one of the \emph{shortest} solutions that covers a given subset of examples.




\smallskip

Powered by just-in-time learning, \tool is able to find the solution \scode{replace-6} 
within 23 seconds, starting from a uniform PCFG:
only a slight slowdown compared with having a biased PCFG from the start.
Note that \euphony, which uses a probabilistic model learned from a corpus of existing solutions,
is unable to solve this benchmark even after 10 minutes.
%


\section{Guided Bottom-up Search}\label{sec:algo}

In this section, we describe our guided bottom-up search algorithm. We first formulate our problem of guided search as an instance of an inductive \sygus problem. We then present our algorithm that enumerates programs in the order of decreasing likelihood. 

\subsection{Preliminaries}

\mypara{Context-free Grammar}
A \emph{context-free grammar} (CFG) is a quadruple $\grammar = (\nontermset, \termset, \start, \ruleset)$, 
where $\nontermset$ denotes a finite, non-empty set of non-terminal symbols, 
$\termset$ denotes a finite set of terminals, 
$\start$ denotes the starting non-terminal, 
and $\ruleset$ is the set of production rules.
In our setting,
each terminal $t \in \termset$ is associated with an \emph{arity} $\iarity(t) \geq 0$,
and each production rule $\rl \in \ruleset$ is of the form $\nterm \to (t\ \nterm_1\ \ldots\ \nterm_k)$, 
where $\nterm,\nterm_1,\ldots,\nterm_k\in\nontermset$, $t\in \termset$, and $\iarity(t) = k$%
\footnote{An astute reader might have noticed that we can formalize this grammar as a \emph{regular tree grammar} instead;
we decided to stick with the more familiar context-free grammar for simplicity.}.
We denote with $\ruleset(\nterm)$ the set of all rules $\rl \in \ruleset$ whose left-hand side is $\nterm$.
A sequence $\alpha \in (\nontermset \cup \termset)^*$ is called a \emph{sentential form}
and a sequence $s \in \termset^*$ is a called a \emph{sentence}.
A grammar $\grammar$ defines a (leftmost) \emph{single-step derivation} relation on sentential forms:
$s \nterm \alpha \step s \beta \alpha$ if $\nterm \to \beta \in \ruleset$.
The reflexive transitive closure of this relation is called (leftmost) \emph{derivation} and written $\manystep$.
All grammars we consider are unambiguous, \ie every sentential form has at most one derivation. 

\mypara{Programs}
A \emph{program} $\prog$ is a sentence derivable from some $\nterm \in \nontermset$;
we call a program \emph{whole} if it is derivable from $\start$.
The set of all programs is called the \emph{language} of the grammar $\grammar$:
$\lang(\grammar) = \{s \in \termset^* \mid \nterm \manystep s\}$.
The \emph{trace} of a program $\trace(\prog)$ is the sequence of production rules 
$\rl_1, \ldots, \rl_n$ used in its derivation ($\nterm \step \alpha_1 \step \ldots \step \alpha_{n-1} \step \prog$).
The \emph{size} of a program $|\prog|$ is the length of its trace.
We assign semantics $\sem{\prog}\colon \vals^*\to \vals$ to each program $\prog$,
where $\vals$ is the set of run-time values.




\mypara{Inductive syntax-guided synthesis}
An \emph{inductive} syntax-guided synthesis (\sygus) problem is defined by a grammar $\grammar$ and 
a set of input-output examples  $\examples = \many{\langle i, o\rangle}$, where $i\in\vals^*$, $o\in \vals$%
\footnote{In general, the $\sygus$ problem allows first-order formulae as a specification, 
and prior work has shown how to reduce this general formulation to inductive formulation using CEGIS~\cite{alur2017scaling, lee2018accelerating}.}.
A \emph{solution} to the problem is a program $P \in \lang(\grammar)$ 
such that $\forall {\langle i, o\rangle} \in \examples$, $\sem{\prog}(i) = o$.
Without loss of generality, we can assume that only whole programs can evaluate to the desired outputs $o$,
hence our formulation need not explicitly require that the solution be whole.

\mypara{Probabilistic Context-free Grammar}
A \emph{probabilistic context-free grammar} (PCFG) $\grammar_p$ 
is a pair of a CFG $\grammar$ and a function $\pcfg : \ruleset \rightarrow [0,1]$ 
that maps each production rule $\rl \in \ruleset$ to its probability. 
Probabilities of all the rules for given non-terminal $\nterm \in \nontermset$ sum up to one: 
$\forall \nterm .\sum_{\rl \in \ruleset(\nterm)} \pcfg(\rl) = 1$.
A PCFG defines a probability distribution on programs:
a probability of a program is the product of probabilities of all the productions in its trace
$\pcfg(\prog) = \prod_{\rl_i\in \trace(\prog)} \pcfg(\rl_i)$.


\mypara{Costs} 
We can define the \emph{real cost} of a production as $\rcost(\rl) = -\log(\pcfg(\rl))$;
then the real costs of a program can be computed as $\rcost(\prog) = -\log(\pcfg(\prog)) = \sum_{\rl_i\in \trace(\prog)} \rcost(\rl_i)$.
For the purpose of our algorithm, we define \emph{discrete costs}, which are real costs rounded to the nearest integer:
$\costfn(\rl) = \round{\rcost(\rl)}$.
The cost of a program $\prog$ is defined as the sum of costs of all the productions in its trace:
$\costfn(\prog) = \sum_{\rl_i\in \trace(\prog)} \costfn(\rl_i)$.

\subsection{Guided Bottom-up Search Algorithm}\label{sec:algo1}

Algorithm~\ref{algo} presents our guided bottom-up search algorithm.
The algorithm takes as input a PCFG $\grammar_p$ and a set of input-output examples $\examples$,
and enumerates programs in the order of increasing discrete costs according to $\grammar_p$,
until it finds a program $\prog$ that satisfies the entire specification $\examples$ 
or reaches a certain cost limit $\llim$.
The algorithm maintains a search state that consists of
\begin{inparaenum}[(1)]
\item the current cost level $\level$;
\item program bank $\bank$, which stores all enumerated programs indexed by their cost;
\item evaluation cache $\exec$, which stores evaluation results of all programs in $\bank$
(for the purpose of checking observational equivalence); and
\item the set $\partialSols$, which stores all enumerated partial solutions.
\end{inparaenum}
Note that the algorithm returns the current search state
and optionally takes a search state as input;
we make use of this in \autoref{sec:update} to resume search from a previously saved state.


\begin{algorithm}[t]
\small
	\caption{Guided Bottom-up search algorithm}\label{algo}
	\begin{algorithmic}[1]
    \Require{PCFG $\grammar_p$, input-output examples  $\examples$, and optionally, the initial state of the search}
    \Ensure{A solution $\prog$ or $\bot$, and the current state of the search}
		\Procedure{Guided-Search}{$\grammar_p,
		\examples, \langle\level_0, \bank_0, \exec_0, \partialSols_0\rangle = \langle0,\emptyset,\emptyset,\emptyset\rangle$}
		\State $\level, \bank, \exec, \partialSols \gets \level_0, \bank_0, \exec_0, \partialSols_0$ 
    \Comment{Initialize state of the search}
		\While{$\level \leq \level_0 + \llim$}
      \For{$\prog \in \textproc{New-Programs}(\grammar_p, \level, \bank)$}  \Comment{For all programs of cost $\level$}
				
		    \State $\eval \gets [{\langle i, \sem{\prog}(i)\rangle} \mid \langle i, o\rangle \in \examples]$
            \Comment{Evaluate on inputs from $\examples$}
		
        \If{($\eval = \examples$)}
          \State \textbf{return} ($\prog, \langle\level, \bank, \exec, \partialSols\rangle$)  \Comment{$\prog$ fully satisfies $\examples$, solution found!}
        \ElsIf{($\eval \in \exec$)}
		      \State{\textbf{continue}}                        \Comment{$\prog$ is observationally equivalent to another program in $\bank$}
        \ElsIf{($\eval \cap \examples \neq \emptyset$)} \Comment{$\prog$ partially satisfies $\examples$}
            \State $\partialSols \gets \partialSols \cup \prog$
        \EndIf    
        \State $\bank[\level] \gets \bank[\level] \cup \{\prog\}$ \Comment{Add to the bank, indexed by cost}
        \State $\exec \gets \exec \cup \eval$                                     \Comment{Cache evaluation result}			  
			\EndFor		
		\State $\level \gets \level + 1$
    \EndWhile
    \State \textbf{return} ($\bot, \langle\level, \bank, \exec, \partialSols\rangle$)    \Comment{Cost limit reached}
    \Statex
		\EndProcedure
    
    \Require{PCFG $\grammar_p$, cost level $\level$, program bank $\bank$ filled up to $\level - 1$}
    \Ensure{Iterator over all programs of cost $\level$}      \Comment{For all production rules}
		\Procedure{New-Programs}{$\grammar_p$, $\level$, $\bank$}
			\For{($\rl = \nterm \rightarrow (t\ \nterm_1\ \nterm_2\ \ldots\ \nterm_k) \in \ruleset$)}				
        \If{$\costfn(\rl) = \level \land k = 0$}  \Comment{$t$ has arity zero}
          \State $\mathbf{yield}\ t$                
        \ElsIf{$\costfn(\rl) < \level \land k > 0$} \Comment{$t$ has non-zero arity}            
          \For{$(c_1,\dots,c_k) \in \set{[1,\level]^k}{\sum c_i = \level - \costfn(\rl)}$}        \Comment{For all subexpression costs}
          \For{$(P_1,\dots,P_k)\in \set{\bank[c_1]\times\ldots\times\bank[c_k]}{\bigwedge_i \nterm_i \manystep P_i}$} \Comment{For all subexpressions}
              \State	$\mathbf{yield}\ (t\ P_1\ \dots\ P_k)$
          \EndFor
          \EndFor
        \EndIf
      \EndFor
    \EndProcedure
	\end{algorithmic}
\end{algorithm}

Every iteration of the loop in lines 3--14 enumerates all programs whose costs are equal to $\level$.
New programs with a given cost are constructed by the auxiliary procedure \textproc{New-Programs},
which we describe below.
In line 5, every new program $P$ is evaluated on the inputs from the semantic specification $\examples$;
if the program matches the specification exactly, it is returned as the solution.
Otherwise, if the evaluation result is already present in $\exec$,
then $P$ is deemed observationally equivalent to another program in $\bank$ and discarded.
A program with new behavior is added to the bank at cost $\level$
and its evaluation result is cached in $\exec$;
moreover, if the program satisfies some of the examples in $\examples$,
it is considered a partial solution and added to $\partialSols$.

The auxiliary procedure \textproc{New-Programs} takes in the PCFG $\grammar_p$,
the current cost $\level$,
and a bank $\bank$ where all levels below the current one are fully filled.
It computes the set of all programs of cost $\level$ in $\grammar_p$.
For the sake of efficiency, instead of returning the whole set at once,
\textproc{New-Programs} is implemented as an \emph{iterator}:
it yields each newly constructed program lazily,
and will not construct the whole set if a solution is found at cost $\level$.
To construct a program of cost $\level$, the procedure iterates over all production rules $\rl \in \ruleset$.
Once $\rl$ is chosen as the top-level production in the derivation of the new program,
we have a budget of $\level - \costfn(\rl)$ to allocate between the subexpressions;
line 21 iterates over all possible subexpression costs that add up to this budget.
Once the subexpression costs $c_1,\ldots,c_k$ have been fixed,
line 22 iterates over all $k$-tuples of programs from the bank that have the right costs and the right \emph{types} to serve as subexpressions:
$\nterm_i \manystep P_i$ means that $P_i$ can replace the nonterminal $\nterm_i$
in the production rule $\rl$.
Finally, line 23 builds a program from the production rule $\rl$ and the subexpressions $P_i$.

\subsection{Guarantees}\label{sec:algo:guarantees}

\mypara{Soundness}
The procedure \textproc{Guided-Search} is \emph{sound}:
given $\grammar_p = \langle\grammar, \pcfg\rangle$ and $\examples$,
if the procedure returns $(\prog, \_)$, then $\prog$ is a solution to the inductive \sygus problem $(\grammar, \examples)$.
It is straightforward to show that $\prog$ satisfies the semantic specification $\examples$,
since we check this property directly in line 6.
Furthermore, $\prog \in \lang(\grammar)$, 
since $\prog$ is constructed by applying a production rule $\rl$ to programs derived from appropriate non-terminals
(see check in line 22).

\mypara{Completeness}
The procedure \textproc{Guided-Search} is \emph{complete}:
if $\prog^*$ is a solution to the inductive \sygus problem $(\grammar, \examples)$,
such that $\costfn(\prog^*) = C$,
and $C \leq \level_0 + \llim$,
then the algorithm will return $(\prog, \_)$, where $\costfn(\prog) \leq C$.
Completeness follows by observing that each level of the bank
is complete up to observational equivalence:
if $\prog\in \lang(\grammar)$ and $\costfn(\prog) \leq C$,
then at the end of the iteration with $\level = C$,
either $\prog\in \bank$ or $\exists \prog'\in \bank\ \suchthat \costfn(\prog') \leq \costfn(\prog)$
and $\forall {\langle i, o\rangle} \in \examples\ \suchthat \sem{\prog}(i) = \sem{\prog'}(i)$.
This in turn follows from the completeness of \textproc{New-Programs}
(it considers all combinations of costs of $\rl$ and the subexpressions that add up to $\level$),
monotonicity of costs
(replacing a subexpression with a more expensive one yields a more expensive program)
and compositionality of program semantics
(replacing a subexpression with an observationally equivalent one yields an observationally equivalent program).

\mypara{Prioritization}
We would also like to claim that \textproc{Guided-Search}
enumerates programs in the order of decreasing likelihood.
This property would hold precisely if we were to enumerate programs in order of increasing real cost $\rcost$:
since the $\log$ function is monotonic, $\pcfg(\prog_1) < \pcfg(\prog_2)$ iff $\rcost(\prog_1) < \rcost(\prog_2)$.
Instead \textproc{Guided-Search} enumerates programs in the order of increasing discrete cost $\costfn$,
so this property only holds approximately due to the rounding error.
Empirical evaluation shows, however, that this approximate prioritization is effective in practice (\autoref{sec:eval}).



\section{Just in time learning}\label{sec:update}

In this section, we introduce a new technique we call just-in-time learning that updates the probabilistic model used to guide synthesis by learning from partial solutions. 
We first present the overall \tool algorithm in \autoref{algo:summary}
and then discuss the three steps involved in updating the PCFG in the remainder of the section. 
%

\begin{algorithm}[t]
\small
	\caption{The \tool algorithm}\label{fig:algo-probe}	
	\begin{algorithmic}[1]
		\Require{CFG $\grammar$, set of input-output examples $\examples$}
		\Ensure{A solution $\prog$ or $\bot$}
		\Procedure{\tool}{$\grammar, \examples$}
      \State $\grammar_p \gets \langle \grammar, p_u \rangle$ \Comment{Initialize PCFG to uniform}
      \State $\level, \bank, \exec \gets 0, \emptyset, \emptyset$ \Comment{Initialize search state}
      \While{not timeout}
      \State $\prog,\langle\level, \bank, \exec, \partialSols\rangle \gets$ \textproc{Guided-Search} ($\grammar_p, \examples, \langle\level, \bank, \exec, \emptyset\rangle$)                                           \Comment{Search with current PCFG $\grammar_p$}
      \If {$\prog \neq \bot$}
        \State \textbf{return} $\prog$                             \Comment{Solution found}
      \EndIf
      \State $\partialSols \gets \textproc{Select}(\partialSols, \exec)$ \Comment{Select promising partial solutions}
      \If {$\partialSols\neq \emptyset$}      
        \State $\grammar_p \gets \textproc{Update}(\grammar_p, \partialSols, \exec)$ \Comment{Update the PCFG $\grammar_p$}
        \State $\level, \bank, \exec \gets 0, \emptyset, \emptyset$           \Comment{Restart the search}
      \EndIf
      \EndWhile
      \State \textbf{return} $\bot$
		\EndProcedure
	\end{algorithmic}
\end{algorithm}

\subsection{Algorithm summary}\label{algo:summary}
The overall structure of the \tool algorithm is presented in Algorithm~\ref{fig:algo-probe}. 
The algorithm iterates between the following two phases until timeout is reached:
\begin{enumerate}
\item \emph{Synthesis phase} searches over the space of programs in order of increasing discrete costs 
using the procedure \textproc{Guided-Search} from \autoref{sec:algo}.
\item \emph{Learning phase} updates the PCFG using the partial solutions found in the synthesis phase.
\end{enumerate}

\tool takes as input an inductive \sygus problem $\grammar, \examples$.
It starts by initializing the PCFG with CFG $\grammar$ and a uniform distribution $\pcfg_u$,
which assigns every production rule $\rl = \nterm \to \beta$ the probability $\pcfg(\rl) = 1/|\ruleset(\nterm)|$.
Each iteration of the \textbf{while}-loop corresponds to one synthesis-learning cycle.
In each cycle, \tool first invokes \textproc{Guided-Search} with the current search state.
If the search finds a solution, \tool terminates successfully (line 7);
otherwise it enters the learning phase, which consists of three steps.
First, procedure \textproc{Select} selects \emph{promising} partial solutions (line 8);
if no such solutions have been found, the search simply resumes from the current state.
Otherwise (line 9),
the second step is to use the promising partial solutions to \textproc{Update} the PCFG,
and the third step is to restart the search (line 11).
These three steps are detailed in the rest of this section.

\subsection{Selecting Promising Partial Solutions}\label{sec:partial}

The procedure \textproc{Select} takes as input the set of partial solutions $\partialSols$
returned by \textproc{Guided-Search},
and selects the ones that are \emph{promising} and should be used to update the PCFG.
%
%
We illustrate this process using the synthesis problem in \autoref{fig:examples-learn};
some partial solutions generated for this problem are listed in \autoref{fig:partial}.
The shortest full solution for this problem is:
\[\scode{(substr arg (- (indexof arg "-" 3) 3) 3)}\]
\mypara{Objectives}
An effective selection procedure must balance the following two objectives.
\begin{asparaenum}[(a)]
\item \emph{Avoid rewarding irrelevant productions:}
The reason we cannot simply use \emph{all} generated partial solutions to update the PCFG
is that partial solutions often contain irrelevant subprograms,
which do not in fact contribute to solving the synthesis problem;
rewarding productions from these irrelevant subprograms derails the search.
For example, consider $P_0$ and $P_1$ in \autoref{fig:partial}:
intuitively, these two programs solve the examples $\{e_0, e_1\}$ \emph{in the same way}, 
but $P_1$ also performs an extraneous character replacement,
which happens to not affect its behavior on these examples.
Hence, we would like to discard $P_1$ from consideration to avoid rewarding the irrelevant production \scode{replace}.
Observe that $P_0$ and $P_1$ satisfy the same subset of examples but $P_1$ has a higher cost;
this suggests discarding partial solutions that are subsumed by a cheaper program.
 

\item \emph{Reward different approaches:}
On the other hand, different partial solutions might represent inherently different approaches to solving the task at hand.
For example, consider partial solutions $P_0$ and $P_2$ in \autoref{fig:partial}; 
intuitively, they represent different strategies for computing the starting position of the substring:
fixed index \vs search (\scode{indexof}).
We would like to consider $P_2$ promising:
indeed, \scode{indexof} turns out to be useful in the final solution.
We observe that although $P_2$ solves the same number of examples and has a higher cost than $P_0$,
it solves a different \emph{subset} of examples, and hence should be considered promising.

\end{asparaenum}

\begin{figure}[t]
	\begin{tabular}{|c|c|c|}
		\hline
		\textbf{ID} &\textbf{Input} & \textbf{Output}  \\ \hline
		$e_0$ & \scode{ "+95 310-537-401"} & \scode{"310"}\\ \hline
		$e_1$ & \scode{ "+72 001-050-856"} & \scode{"001"} \\ \hline	
		$e_2$ & \scode{ "+106 769-858-438"} & \scode{"769"} \\ \hline		
	\end{tabular}
	\caption{A set of input-output examples for a string transformation (adapted from \cite{euphony}).}\label{fig:examples-learn}
\end{figure}


\begin{figure}[t]
	\begin{tabular}{|c|c|c|c|c|}
		\hline
		\textbf{Cycle} &\textbf{ID} &\textbf{Examples Satisfied} & \textbf{Partial Solutions} & \textbf{Cost} \\ \hline
		1 & $P_0$ & $\{e_0, e_1\}$ & \scode{(substr arg 4 3)} & 20 \\ \hline
	    2 & $P_1$ & $\{e_0, e_1\}$ & \scode{(replace (substr arg 4 3) " " arg)} & 21 \\ \hline    
		3 & $P_2$ & $\{e_1, e_2\}$ & \scode{(substr arg (indexof arg (at arg 5) 3) 3)} & 37\\ \hline
		3 & $P_3$ & $\{e_1, e_2\}$ & \scode{(substr arg (- 4 (to.int (at arg 4))) 3)} & 37\\ \hline
	\end{tabular}
	\caption{Partial solutions and the corresponding subset of examples satisfied for the problem in \autoref{fig:examples-learn}}\label{fig:partial}
\end{figure}


Our goal is to find the right trade-off between the two objectives.
Selecting too many partial solutions might lead to rewarding irrelevant productions
and more frequent restarts 
(recall that search is restarted only if new promising partial solutions were found in the current cycle).
On the other hand,
selecting too few partial solutions might lead the synthesizer down the wrong path 
or simply not provide enough guidance, especially when the grammar is large.
%
%

\mypara{Selection schemes}
Based on these objectives, we designed three \emph{selection schemes},
which make different trade-offs and are described below from most to least selective.
Note that all selection schemes need to preserve information about promising partial solutions  
between different synthesis-learning cycles,
to avoid rewarding the same solution again after synthesis restarts.
We evaluate the effectiveness of these schemes in comparison to the baseline (using all partial solutions)
in \autoref{sec:eval}.
\begin{asparaenum}
\item \textsc{Largest Subset}:
This scheme selects \emph{a single cheapest} program (first enumerated) 
that satisfies \emph{the largest subset} of examples encountered so far across all synthesis cycles.
Consequently, the number of promising partial solutions it selects is always smaller than the size of $\examples$.
Among partial solutions in \autoref{fig:partial}, this scheme picks a single program $P_0$.



\item \textsc{First Cheapest}:
This scheme selects \emph{a single cheapest} program (first enumerated) 
that satisfies \emph{a unique subset} of examples.
The partial solutions $\{P_0, P_2\}$ from \autoref{fig:partial} are selected by this scheme.
This scheme still rewards a small number of partial solutions, but allows different approaches to be considered. 

\item \textsc{All Cheapest}:
This scheme selects \emph{all cheapest} programs (enumerated during a single cycle)
that satisfy \emph{a unique subset} of examples.
The partial solutions $\{P_0, P_2, P_3\}$ are selected by this scheme. 
Specifically, $P_2$ and $P_3$ satisfy the same subset of examples; 
both are considered since they have the same cost. 
This scheme considers more partial solutions than \textsc{First Cheapest}, which refines the ability to reward different approaches.

%
\end{asparaenum}

\subsection{Updating the PCFG}\label{sec:rule}

Procedure \textproc{Update} uses the set of promising partial solution $\partialSols$
to compute the new probability for each production rule $\rl\in \ruleset$ using the formula:
\[
	\pcfg(\rl) = \dfrac{\pcfg_{u}(\rl) ^ {(1 - \fit)}}{Z}
  \quad\quad\text{where}\quad\quad \fit=\max_{\{P\in \partialSols \mid \rl \in \trace(P)\}} \dfrac{|\examples \cap \exec[P]|}{|\examples|}
	\label{updaterule}
\]
where $Z$ denotes the normalization factor, 
and \fit~ is the highest proportion of input-output examples
that any partial solution derived using this rule satisfies.
Recall that $\pcfg_{u}$ is the uniform distribution for $\grammar$.
This rule assigns higher probabilities to rules that occur in partial solutions that satisfy many input-output examples.



\subsection{Restarting the Search}

Every time the PCFG is updated during a learning phase,
\tool restarts the bottom-up enumeration from scratch,
\ie empties the bank $\bank$ (and the evaluation cache $\exec$)
and resets the current cost $\level$ to zero.
At a first glance this seems like a waste of computation:
why not just resume the enumeration from the current state?
The challenge is that any update to the PCFG renders the program bank outdated,
and updating the bank to match the new PCFG requires the amount of computation and/or memory
that does not pay off in relation to the simpler approach of restarting the search.
Let us illustrate these design trade-offs with an example.

Consider again the synthesis problem in \autoref{fig:examples-learn},
and two programs encountered during the first synthesis cycle:
the program \scode{0} with cost 5 and the program \scode{(indexof arg "+")} with cost 15.
Note that both programs evaluate to 0 on all three example inputs,
\ie they belong to the same observational \emph{equivalence class} $[0,0,0]$;
hence the latter program is \emph{discarded} by observational equivalence reduction,
while the former, discovered first, is chosen as the \emph{representative} of its equivalence class and appears in the current bank $\bank$.

Now assume that during the subsequent learning phase the PCFG changed in such a way
that the new costs of these two programs are $\costfn(\scode{0}) = 10$ and $\costfn(\scode{(indexof arg "+")}) = 7$.
Let us examine different options for the subsequent synthesis cycle. 

\begin{asparaenum}[(1)]
\item \emph{Restart from scratch:}
If we restart the search with an empty bank,
the program \scode{(indexof arg "+")} is now encountered before the program \scode{0}
and selected as the representative of it equivalence class.
In other words, the desired the behavior under the new PCFG
is that the class $[0,0,0]$ has cost 7.
Can we achieve this behavior without restarting the search?

\item \emph{Keep the bank unchanged:} 
Resuming the enumeration with $\bank$ unchanged would be incorrect:
in this case the representative of $[0,0,0]$ is still the program \scode{0} with cost 5.
As a result, any program we build in the new cycle 
that uses this equivalence class as a sub-program would have a wrong cost,
and hence the enumeration order would be different from that prescribed by the new PCFG.

\item \emph{Re-index the bank:}
Another option is to keep the programs stored in $\bank$ but re-index it with their updated costs:
for example, index the program \scode{0} with cost 10.
This does not solve the problem, however:
now class $[0,0,0]$ has cost 10 instead the desired cost 7,
because it still has a wrong representative in $\bank$.
Therefore, in order to enforce the correct enumeration order in the new cycle
we need to update the equivalence class representatives stored in the bank.

\item \emph{Update representatives:}
To be able to update the representatives, we need to store the redundant programs in the bank
instead of discarding them.
To this end, prior work~\cite{bodik2016superoptimizaiton,Wang2017FiniteTreeAutomata,Wang2017AbstractionRefinement}
has proposed representing the bank as a \emph{finite tree automaton},
\ie a hypergraph where nodes correspond to equivalence classes (such as $[0,0,0]$)
and edges correspond to productions (with the corresponding arity).
The representative program of an equivalence class can be computed 
as the shortest hyper-path to the corresponding node 
from the set of initial nodes (inputs and literals);
the cost of the class is the length of such a shortest path.
When the PCFG is updated, leading to modified costs of hyper-edges,
shortest paths for all nodes in this graph need to be recomputed.
Algorithms for doing so~\cite{hypergraphs} have super-linear complexity in the number of affected nodes.
Since in our case most nodes are likely to be affected by the update,
and since the number of nodes in the hypergraph is the same as the size of our bank $\bank$,
this update step is roughly as expensive as rebuilding the bank from scratch.
In addition, for a search space as large as the one \tool explores for the \sygus String benchmarks,
the memory overhead of storing the entire hypergraph is also prohibitive.
\end{asparaenum}

\smallskip
Since restarting the search is expensive, 
\tool does not return from the guided search immediately once a partial solution is found
and instead keeps searching until a fixed cost limit
and returns partial solutions in batches. 
There is a trade-off between restarting synthesis too often 
(wasting time exploring small programs again and again) 
and restarting too infrequently
(wasting time on unpromising parts of the search space when an updated PCFG could guide the search better).
In our implementation, we found that setting the cost limit to $6\cdot C$ works best empirically,
where $C$ is the maximum production cost in the initial PCFG
(this roughly corresponds to enumerating programs in size increments of six with the initial grammar).

\section{Experiments}\label{sec:eval}

We have implemented the \tool synthesis algorithm in Scala%
\footnote{\url{https://github.com/shraddhabarke/probe.git}}. 
In this section, we empirically evaluate how \tool compares to the baseline and state-of-the-art synthesis techniques. 
We design our experiments to answer the following research questions:

\begin{enumerate}[label=(\bfseries Q\arabic*)]
	\item
	How effective is the just-in-time learning in \tool? We examine this question in two parts:
	\begin{enumerate}[label=\arabic*.]
	\item by comparing \tool to unguided bottom-up enumerative techniques, and
	\item by comparing different schemes for partial solution selection.
	\end{enumerate}
	\item 
	Is \tool faster than state-of-the-art \sygus solvers?\label{rq-faster}
	\item
	Is the quality of \tool solutions comparable with state-of-the-art \sygus solvers?\label{rq-quality}
\end{enumerate}

\subsection{Experimental Setup}


We evaluate \tool on three different application domains: 
string manipulation (\stringbench), 
bit-vector manipulation (\bitvec), 
and circuit transformations (\circuit).
We perform our experiments on a set of total 140 benchmarks, 82 of which are \stringbench benchmarks, 27 are \bitvec benchmarks and 31 are \circuit benchmarks.
The grammars containing the available operations for each of these domains appear 
\iflong
in the appendix.
\else
in the extended version of this paper~\cite{extended}.
\fi

\mypara{\stringbench Benchmarks}
The 82 \stringbench benchmarks are taken from the testing set of \euphony~\cite{euphony}. The entire \euphony String benchmark suite consists of 205 problems, from the PBE-String track of the 2017 \sygus competition and from string-manipulation questions from popular online forums. 
\euphony uses 82 out of these 205 benchmarks as their testing set based on the criterion that \eusolver~\cite{alur2017scaling} could not solve them within 10 minutes. 
\stringbench benchmark grammars have a median of 16 operations, 11 literals, and 1 variable.
All these benchmarks use input-output examples as semantic specification,
and the number of examples ranges from 2 to 400.

\mypara{\bitvec Benchmarks} The 27 \bitvec benchmarks originate from the book \emph{Hacker's Delight}~\cite{warren2013hacker},
commonly referred to as the bible of bit-twiddling hacks.
We took 20 of them verbatim from the \sygus competition suite:
these are all the highest difficulty level (d5) Hacker's Delight benchmarks in \sygus.
We then found 7 additional loop-free benchmarks in synthesis literature~\cite{jha2010oracle,gulwani2011synthesis}
and manually encoded them in the \sygus format.
%
\bitvec benchmark grammars have a median of 17 operations, 3 literals, and 1 variable.
The semantic specification of \bitvec benchmarks is a universally-quantified first-order formula that is functionally equivalent to the target program.

Note that in addition to Hacker's Delight benchmarks, 
the \sygus bitvector benchmark set also contains \euphony bitvector benchmarks.
We decided to exclude these benchmarks from our evaluation 
because they have very peculiar solutions:
they all require extensive case-splitting, and hence are particularly suited to synthesizers 
that perform \emph{condition abduction}~\cite{alur2017scaling,Kneuss2013synthesis,albarghouthi2013recursive}.
Since \tool (unlike \euphony) does not implement condition abduction, 
it is bound to perform poorly on these benchmarks.
At the same time, condition abduction is orthogonal to the techniques introduced in this paper;
hence \tool's performance on these benchmarks would not be informative.


\mypara{\circuit Benchmarks}
The 31 \circuit benchmarks are taken from the \euphony testing set. These benchmarks involve synthesizing constant-time circuits that are cryptographically resilient to timing attacks.
\circuit benchmark grammars have a median of 4 operations, 0 literals, and 6 variables.
The semantic specification is a universally-quantified boolean formula functionally equivalent to the circuit to be synthesized.

\mypara{Reducing first-order specifications to examples}
As discussed above, only the string domain uses input-output examples as the semantic specification, 
while the other two domains use a more general \sygus formulation where the specification is a (universally-quantified) first-order formula. 
We extend \tool to handle the latter kind of specifications in a standard way (see \eg~\cite{alur2017scaling}), 
using \emph{counter-example guided inductive synthesis} (CEGIS)~\cite{solar2006combinatorial}. 
CEGIS proceeds in iterations, where each iteration first \emph{synthesizes} a candidate program 
that works on a finite set of inputs,
and then \emph{verifies} this candidate against the full specification,
adding any failing inputs to the set of inputs to be considered in the next synthesis iteration.
We use \tool for the synthesis phase of the CEGIS loop.
At the start of each CEGIS iteration, we initialize an independent instance of \tool starting from a uniform grammar. 
%

\mypara{Baseline Solvers} 
As the state-of-the-art in research questions \ref{rq-faster} and \ref{rq-quality} we use \euphony and \cvc, which are
the state-of-the-art \sygus solvers in terms of performance and solution quality. 
\euphony~\cite{lee2018accelerating} also uses probabilistic models to guide its search, but unlike \tool they are pre-learned models. We used the trained models that are available in \euphony's repository \cite{euphony}.
\cvc~\cite{reynolds2019cvc} has been the winner of the PBE-Strings track of the \sygus Competition~\cite{alur2017sygus} since 2017. We use the \cvc version 1.8 (Aug  6 2020 build).

\mypara{Experimental setup}
All experiments were run with a 10 minute timeout for all solvers,
on a commodity Lenovo laptop with a i7 quad-core CPU @ 1.90GHz with 16GB of RAM.

\begin{figure}[h]
	\centering
	\begin{subfigure}{.51\textwidth}
		\centering
		\includegraphics[width=\textwidth]{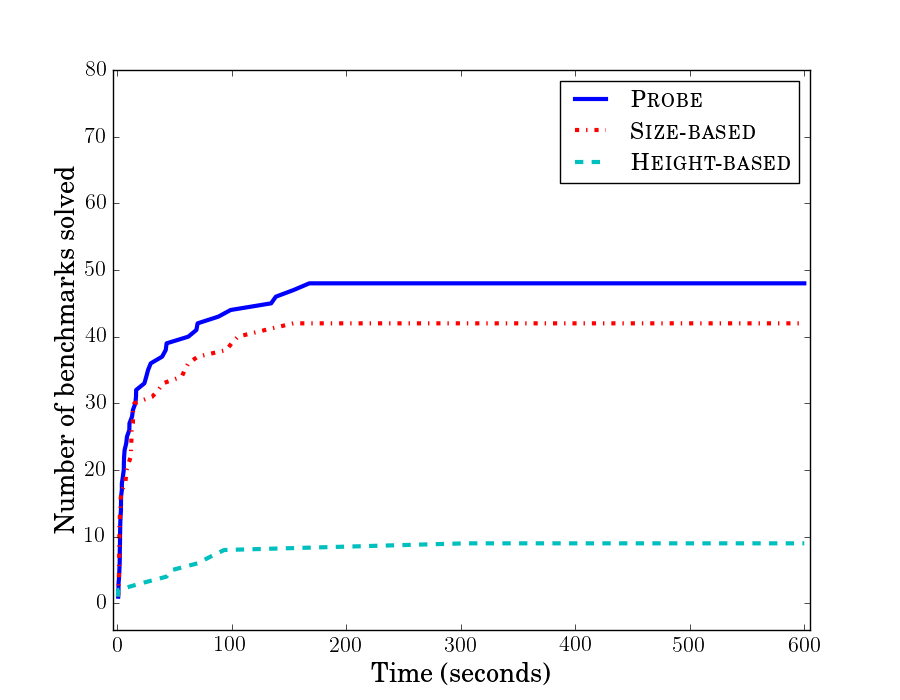}
		\caption{\stringbench domain with regular grammar.}
		\label{fig:string-baseline}
	\end{subfigure}%
	\begin{subfigure}{.51\textwidth}
		\centering
		\includegraphics[width=\textwidth]{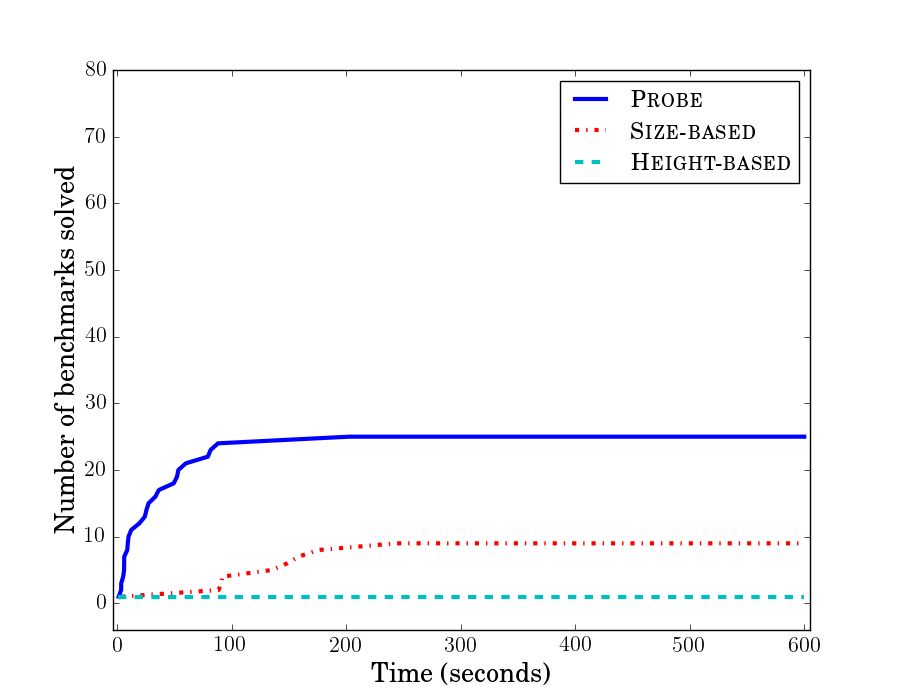}
		\caption{\stringbench domain with extended grammar.}
		\label{fig:string-ext-baseline}
	\end{subfigure}%
	
	\begin{subfigure}{.51\textwidth}
		\centering
		\includegraphics[width=\textwidth]{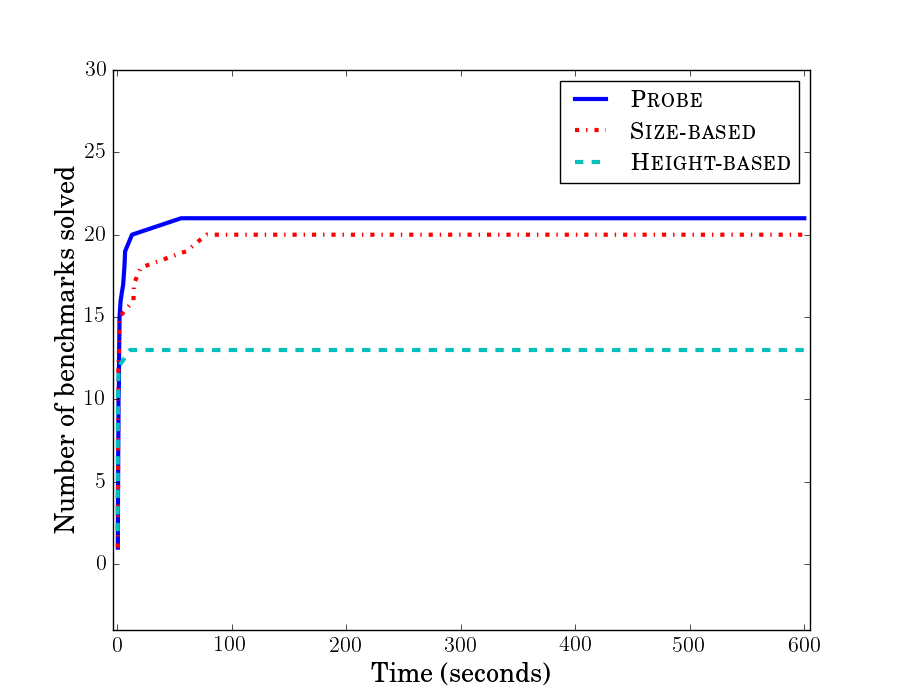}
		\caption{\bitvec domain}
		\label{fig:bitvec-baseline}
	\end{subfigure}%
	\begin{subfigure}{.51\textwidth}
		\centering
		\includegraphics[width=\textwidth]{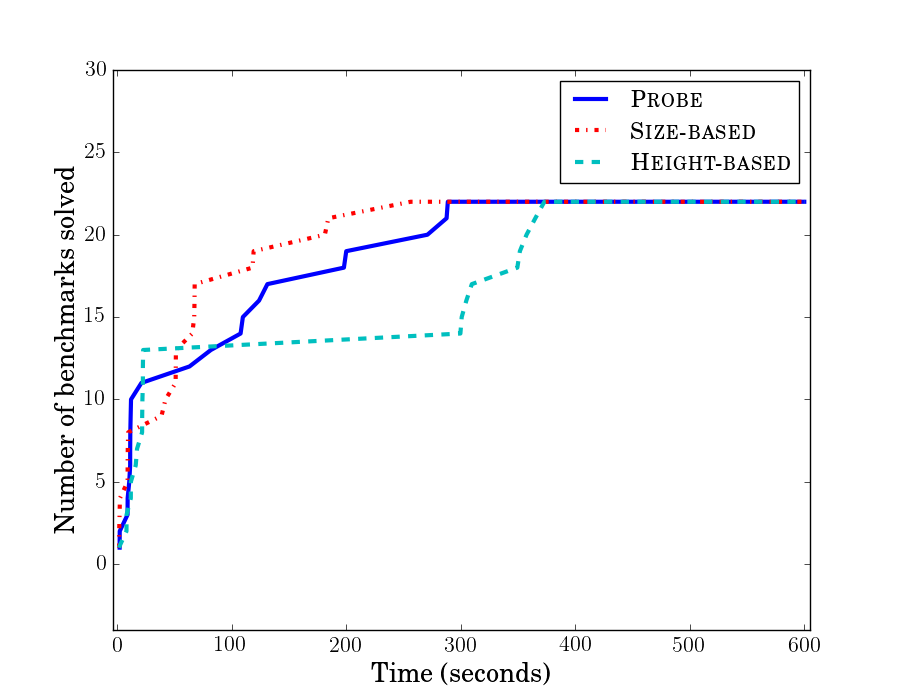}
		\caption{\circuit domain}
		\label{fig:circuit-baseline}
	\end{subfigure}%
	\caption{Number of benchmarks solved by \tool and unguided search techniques (size-based and height-based enumeration) for \stringbench, \bitvec and \circuit domains. 
  Timeout is 10 min, graph scale is linear.}
	\label{fig:probe-vs-unguided-string}
\end{figure}

\subsection{Q1.1: Effectiveness of Just-in-time learning} \label{sec:probe-vs-unguided}

To assess the effectiveness of the  just-in-time learning approach implemented in \tool, 
we first compare it to two unguided bottom-up search algorithms:
height-based and size-based enumeration. 
We implement these baselines inside \tool, as simplifications of guided bottom-up search.

\mypara{Results for \stringbench Domain}
We measure the time to solution for each of the 82 benchmarks in the \stringbench benchmark set, for each of the three methods: 
\tool, size-based, and height-based enumeration.
The results are shown in \autoref{fig:string-baseline}.
\tool, size-based and height-based enumeration are able to solve 48, 42 and 9 problems, respectively. 
Additionally, at every point after one second, \tool has solved more benchmarks than either size-based or height-based enumeration.

\mypara{Just-in-time learning and grammar size}
In addition to our regular benchmark suite, 
we created a version of the \stringbench benchmarks (except 12 outliers that have abnormally many string literals) 
that uses an \emph{extended string grammar}, 
which includes all operations and literals from all \stringbench benchmarks. 
In total this grammar has all available string, integer and boolean operations in the \sygus language specification and 48 string literals and 11 integer literals.
These 70 extended-grammar benchmarks allow us to test the behavior of \tool on larger grammars and thereby larger program spaces.

Within a timeout of 10 minutes, \tool solves 25 benchmarks (52\% of the original number) 
whereas height-based and size-based enumeration solved 1 (11\% of original) and 9 (21\% of original) benchmarks respectively as shown in \autoref{fig:string-ext-baseline}.
We find this particularly encouraging, because the size of the grammar usually has a severe effect on the synthesizer 
(as we can see for size-based enumeration), 
so much so that carefully constructing a grammar is considered to be part of synthesizer design.
%
While the baseline synthesizers need the benefit of approaching each task with a different, carefully chosen grammar, 
\tool's just-in-time learning is much more robust to additional useless grammar productions.
Even with a larger grammar, \tool's search space does not grow as much: once it finds a partial solution, it hones in on the useful parts of the grammar. 

\mypara{Results for \bitvec Domain}
The results for the \bitvec benchmarks are shown in \autoref{fig:bitvec-baseline}.
Out of the 27 \bitvec benchmarks, \tool, size-based and height-based solve 21, 20 and 13 benchmarks, respectively.
In addition to solving one more benchmark,
\tool is also considerably faster than size-based enumeration,
as we can see from the horizontal distance between the two curves on the graph.
%
\tool significantly outperforms the baseline height-based enumeration technique. 

\mypara{Results for \circuit Domain} 
The results for the \circuit benchmarks are shown in \autoref{fig:circuit-baseline}.
Each of the three techniques solves 22 out of 31 benchmarks,
with size-based enumeration outperforming \tool in terms of synthesis times.
%
The reason \tool performs worse in this domain
is that the \circuit grammar is very small (only four operations in the median case) 
and the solutions tend to use most of productions from the grammar. 
Thus, rewarding specific productions in the PCFG does not yield significant benefits, 
but in fact the search is slowed down due to the restarting overhead incurred by \tool.
 
\mypara{Summary of results}
Out of the 210 benchmarks from three different domains and the extended \stringbench grammar, \tool solves 116, size-based solves 93 and height-based solves 45. We conclude that overall, \textbf{\tool outperforms both baseline techniques, and is therefore an effective synthesis technique.}

\subsection{Q1.2: Selection of partial solutions}
\begin{figure}[t]
	\centering
	\begin{subfigure}{.51\textwidth}
		\centering
		\includegraphics[width=\textwidth]{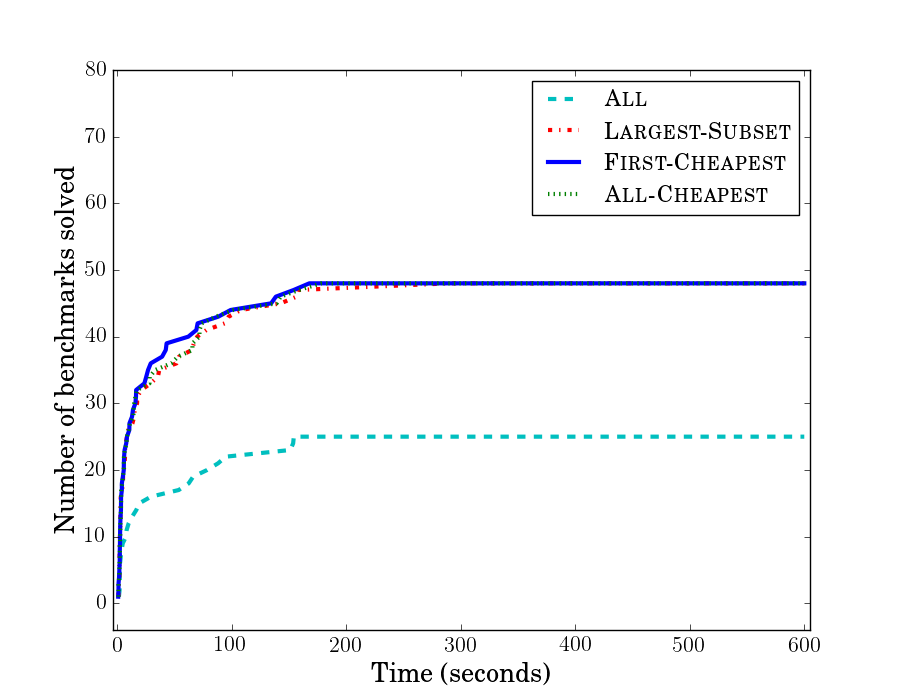}
		\caption{\stringbench domain}
		\label{fig:ablation-string}
	\end{subfigure}%
	
	\begin{subfigure}{.51\textwidth}
		\centering
		\includegraphics[width=\textwidth]{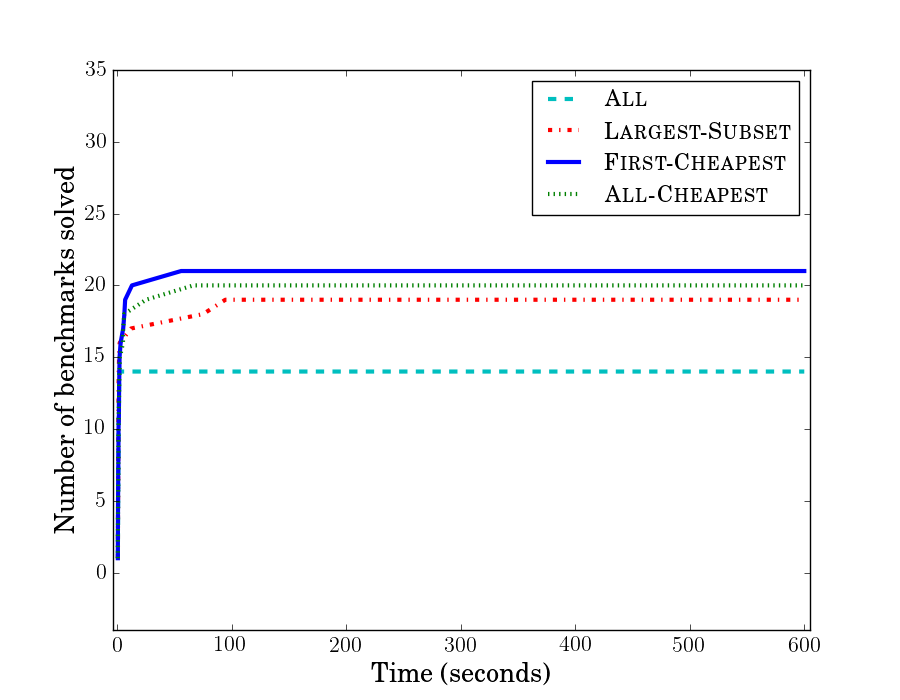}
		\caption{\bitvec domain}
		\label{fig:ablation-bitvec}
	\end{subfigure}%
	\begin{subfigure}{.51\textwidth}
		\centering
		\includegraphics[width=\textwidth]{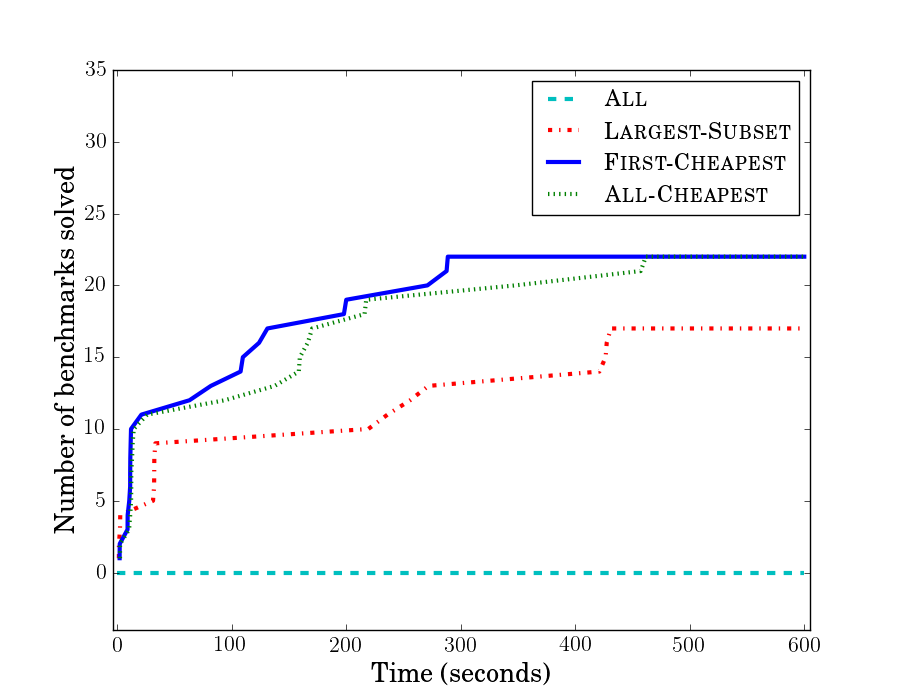}
		\caption{\circuit domain}
		\label{fig:ablation-circuit}
	\end{subfigure}%
	\caption{Number of benchmarks solved by \tool with schemes for selecting promising partial solutions. 
  Schemes are described in \autoref{sec:partial}; \all represents no selection (all partial solutions are used to update the PCFG).
  Timeout is 10 min, graph scale is linear.}
	\label{fig:ablation-study}
\end{figure}

In this section, we empirically evaluate the schemes for selecting promising partial solutions. 
We compare four different schemes: the three described in \autoref{sec:partial}
and the baseline of using \all generated partial solutions.
The results are shown in \autoref{fig:ablation-study}.

The \all baseline scheme performs consistently worse than the other schemes on all three domains
(and also worse than unguided size-based enumeration).
For the circuit domain (\autoref{fig:ablation-circuit}), the \all scheme solves none of the benchmarks.
The performance of the remaining schemes is very similar, 
indicating that the general idea of leveraging small and semantically unique partial solutions to guide search 
is robust to minor changes in the selection criteria.
We select \textsc{First Cheapest} as the scheme used in \tool since it provides a balance between rewarding 
few partial solutions while still considering syntactically different approaches. 
%


\subsection{Q2: Is \tool faster than the state of the art?} \label{sec:probe-vs-euphony}

\begin{figure}[t]
	\centering
	\begin{subfigure}{.51\textwidth}
		\centering
		\includegraphics[width=\textwidth]{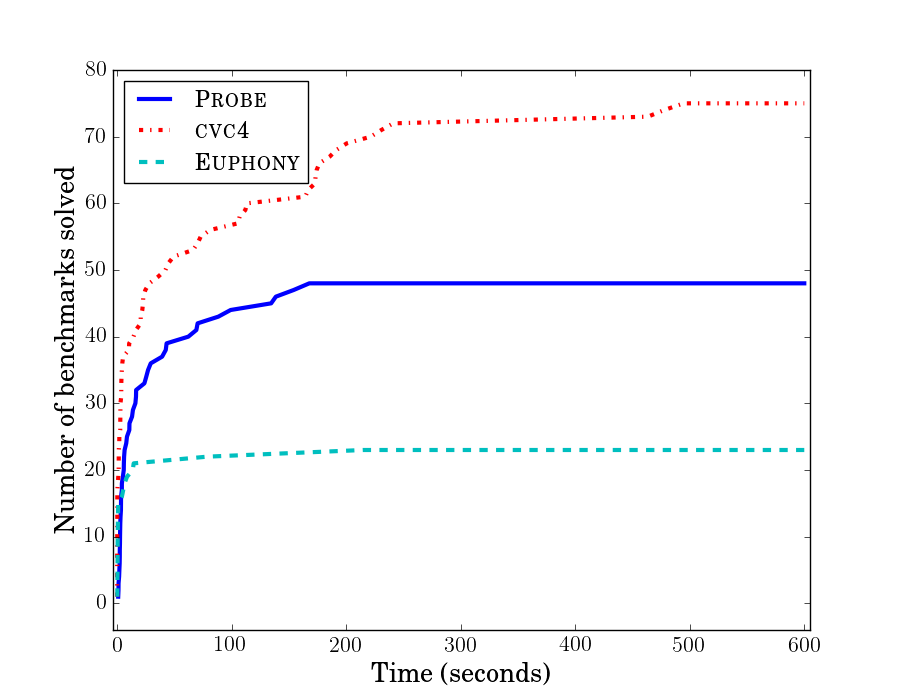}
		\caption{\stringbench domain}
		\label{fig:others-string}
	\end{subfigure}%

	\begin{subfigure}{.51\textwidth}
		\centering
		\includegraphics[width=\textwidth]{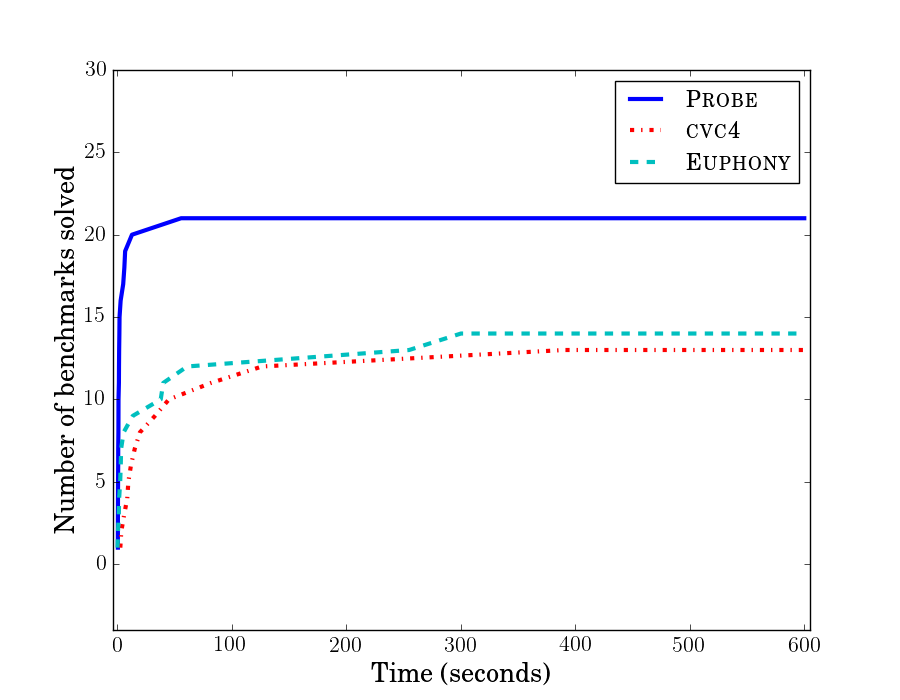}
		\caption{\bitvec domain}
		\label{fig:others-bitvec}
	\end{subfigure}%
	\begin{subfigure}{.51\textwidth}
		\centering
		\includegraphics[width=\textwidth]{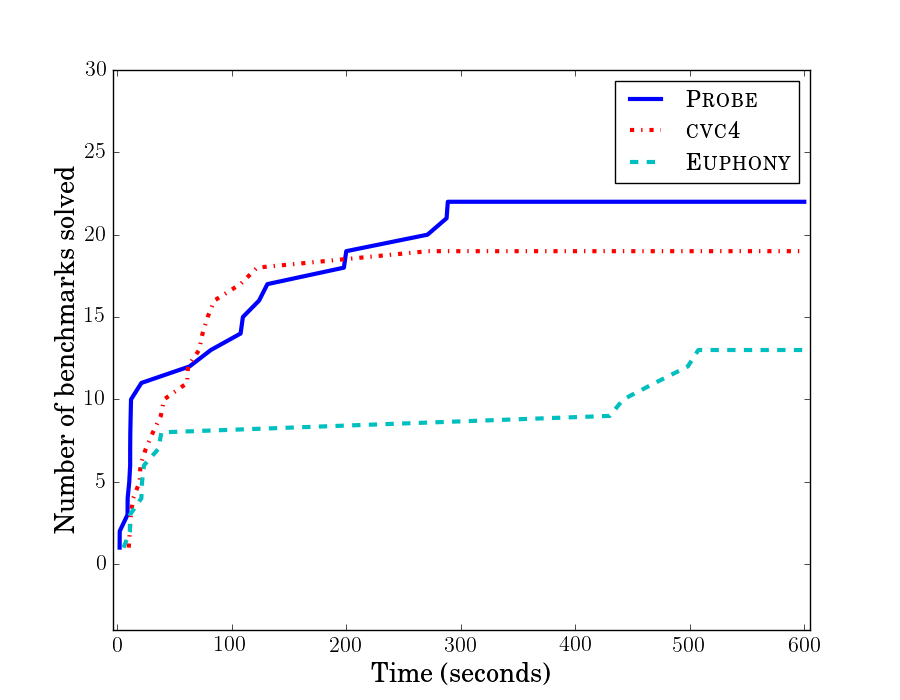}
		\caption{\circuit domain}
		\label{fig:others-circuit}
	\end{subfigure}%
	\caption{Number of benchmarks solved by \tool, \euphony and \cvc for \stringbench, \bitvec and \circuit domains. Timeout is 10 min, graph scale is linear.}
	\label{fig:probe-vs-others}
\end{figure}

We compare \tool's time to solution on the benchmarks in our suite against two state-of-the-art \sygus solvers, \euphony and \cvc.
The results for all three domains are shown in \autoref{fig:probe-vs-others}.

\mypara{\stringbench Domain}
Results for the \stringbench domain are shown in \autoref{fig:others-string}. 
Of the 82 benchmarks in the \stringbench suite, 
\tool solves 48 benchmarks, with an average time of 29s and a median time of 8.3s.
\euphony solves 23 benchmarks, with average of 15.4s and a median of 0.7s. 
\cvc solves 75 benchmarks, with an average of 61.8s and a median of 10.2s.

The performance of \euphony is close to that reported originally by \citet{lee2018accelerating}; they report 27 of the 82 benchmarks solved with a 60 minute timeout.
Even with the reduced timeout, \tool vastly outperforms \euphony.

When only examining time to solution, \cvc outperforms \tool: 
not only does it solve more benchmarks faster, but it still solves new benchmarks long after \tool and \euphony have plateaued.
However, these solutions are not necessarily usable, as we show in \autoref{sec:probe-vs-cvc4}.

\mypara{\bitvec Domain}
Out of the 27 \bitvec benchmarks, \tool solves 21 benchmarks, \euphony solves 14 and \cvc solves 13 benchmarks as shown in \autoref{fig:others-bitvec}. 
\tool outperforms both \cvc and \euphony on these benchmarks with an average time of 5s and median time of 1.5s. 
\euphony's average time is 52s and median is 4.6s while \cvc takes an average of 58s and a median of 15s.
\tool not only solves the most benchmarks overall, it also solves the highest number of benchmarks compared to \euphony and \cvc at each point in time. 

We should note that the \euphony model we used for this experiment
was trained on the \euphony set of bit-vector benchmarks
(the ones we excluded because of the case-splits) rather than the Hacker's Delight benchmarks.
Although \euphony does very well on its own bit-vector benchmarks,
it does not fare so well on Hacker's Delight. 
These results shed some light on how brittle pre-trained models are in the face of subtle changes in syntactic program properties,
even within a single bit-vector domain;
we believe this makes a case for just-in-time learning.


\mypara{\circuit Domain}
Out of the 31 \circuit benchmarks, \tool solves 22 benchmarks with an average time of 90s and median time of 42s
(see \autoref{fig:others-circuit}).  
\euphony solves 13 benchmarks with average and median times of 193.6s and 36s.
\cvc solves 19 benchmarks with average and median times of 60s and 41s. 
\tool outperforms both \cvc and \euphony in terms of the number of benchmarks solved. 
Moreover \cvc generates much larger solutions than \tool, as discussed in \autoref{sec:probe-vs-cvc4}.

\mypara{Summary of results}
Of the total 140 benchmarks, \tool solves 91 within the 10-minute timeout, \euphony solves 50, and \cvc solves 107.
\tool outperforms \euphony's pre-learned models in all three domains, 
and while \cvc outperforms \tool in the \stringbench domain; 
the next subsection will discuss the quality of the results it generates.

\subsection{Q3: Quality of synthesized solutions} \label{sec:probe-vs-cvc4}

So far, we have tested the ability of solvers to arrive at \emph{a} solution, without checking what the solution is.
When a PBE synthesizer finds a program for a given set of examples, 
it guarantees nothing but the behavior on those examples. 
%
Indeed, the \sygus Competition scoring system\footnote{\url{https://sygus.org/comp/2019/results-slides.pdf}, slide 13} awards the most points (five) 
for simply returning any program that matches the given examples.
It is therefore useful to examine the \emph{quality} of the solutions generated by \tool and its competition.

Size is a common surrogate measure for program \emph{simplicity}: 
e.g., the \sygus Competition awards an additional point to the solver that returns the smallest program for each benchmark. 
Program size reflects two sources of complexity:
\begin{inparaenum}[(i)]
\item unnecessary operations that do not influence the result, and, perhaps more importantly,
\item \emph{case splitting} that overfits to the examples.
\end{inparaenum}
It is therefore reasonable to assume that a smaller solution is more interpretable
and generalizes better to additional inputs beyond the initial input-output examples.

Based on these observations, we first estimate the quality of results for all three domains
by comparing the sizes of solutions generated by \tool and other tools.
We next focus on the \stringbench benchmarks,
as this is the only domain where the specification is given in the form of input-output examples,
and hence is prone to overfitting.
For this domain, we additionally measure the number of case splits in generated solutions
and test their generalization accuracy on unseen inputs.



\begin{figure}[t]
	\centering
	\begin{subfigure}{.33\textwidth}
		\centering
		\includegraphics[width=1.05\textwidth]{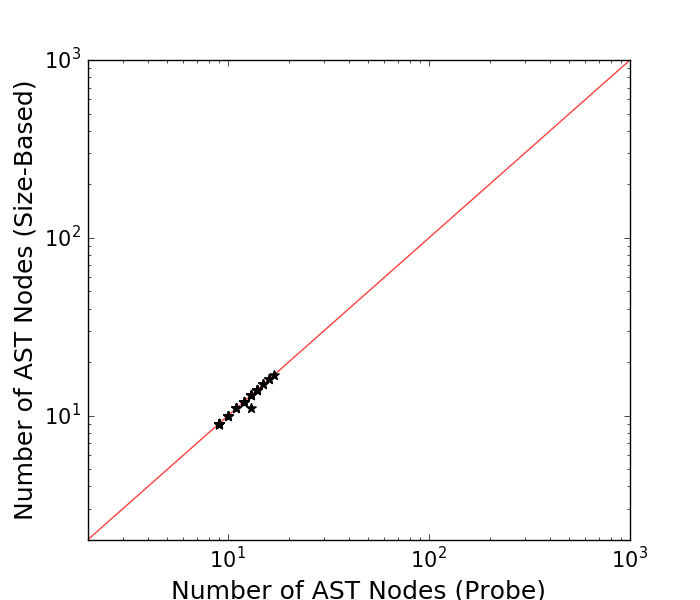}
		\caption{Size-based \stringbench domain}
		\label{fig:sizebased-string}
	\end{subfigure}%
	\begin{subfigure}{.33\textwidth}
		\centering
		\includegraphics[width=1.05\textwidth]{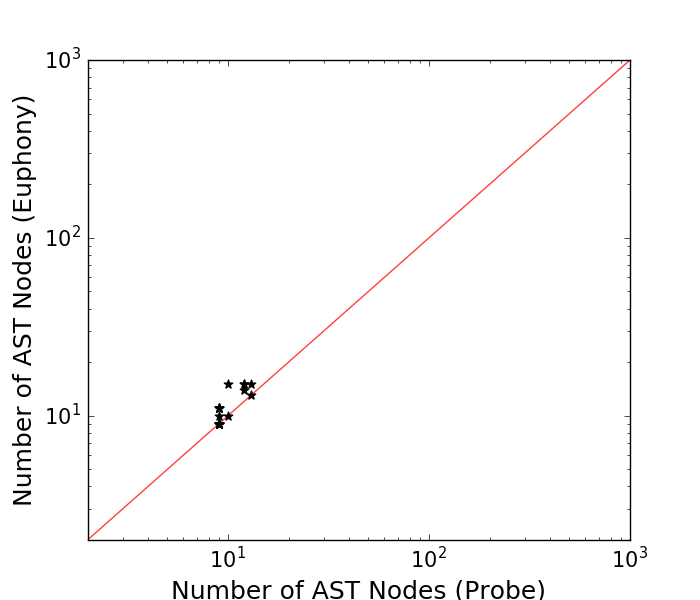}
		\caption{\euphony \stringbench domain}
		\label{fig:euphony-string}
	\end{subfigure}%
	\begin{subfigure}{.33\textwidth}
		\centering
		\includegraphics[width=1.05\textwidth]{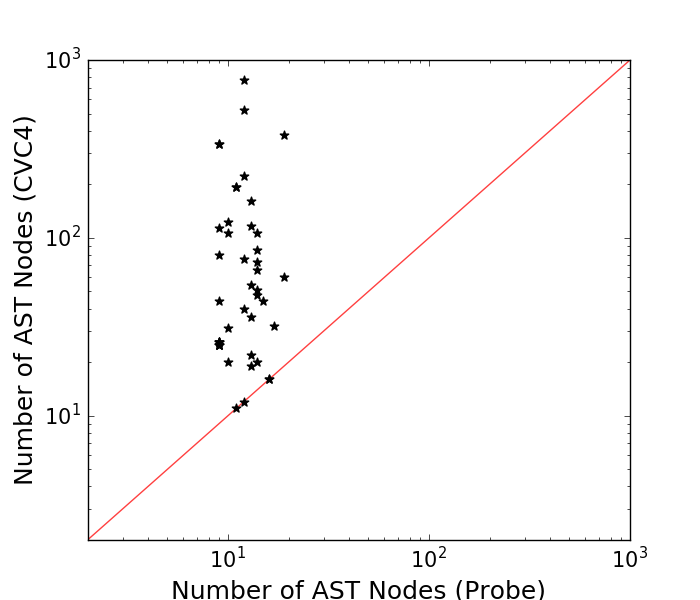}
		\caption{\cvc \stringbench domain}
		\label{fig:cvc4-string}
	\end{subfigure}
	
	\begin{subfigure}{.33\textwidth}
		\centering
		\includegraphics[width=1.05\textwidth]{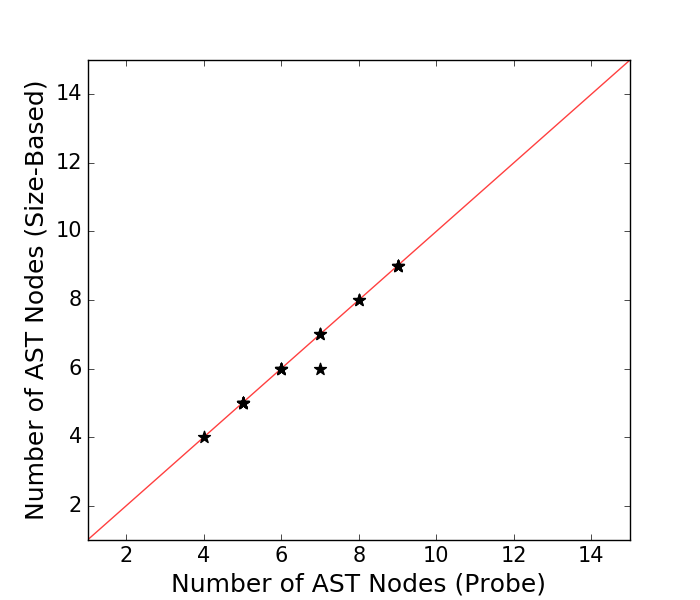}
		\caption{Size-based \bitvec domain}
		\label{fig:sizebased-bitvec}
	\end{subfigure}%
	\begin{subfigure}{.33\textwidth}
		\centering
		\includegraphics[width=1.05\textwidth]{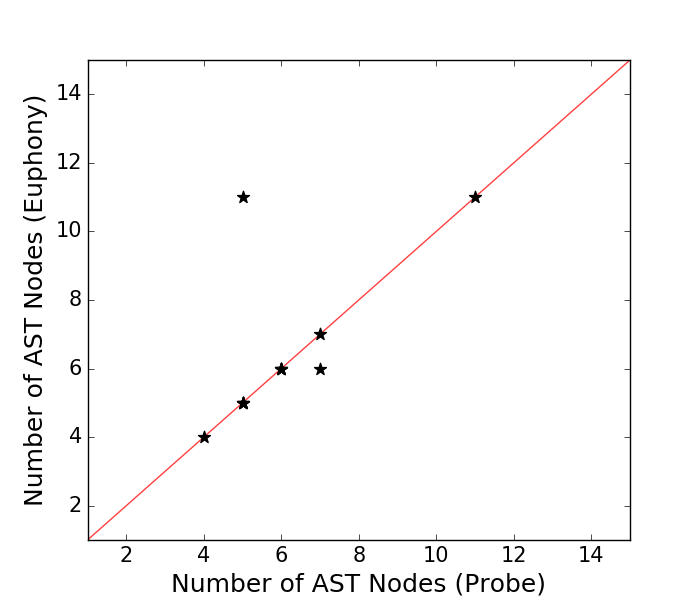}
		\caption{\euphony \bitvec domain}
		\label{fig:euphony-bitvec}
	\end{subfigure}%
	\begin{subfigure}{.33\textwidth}
		\centering
		\includegraphics[width=1.05\textwidth]{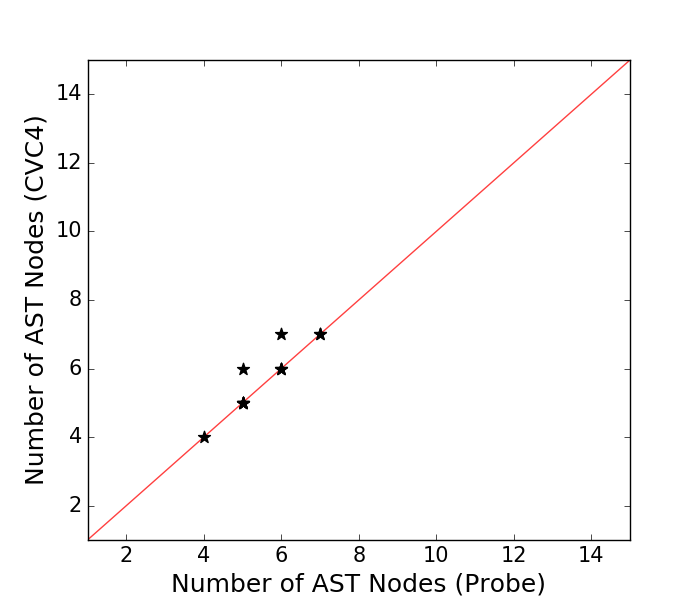}
		\caption{\cvc \bitvec domain}
		\label{fig:cvc4-bitvec}
	\end{subfigure}
	
	\begin{subfigure}{.33\textwidth}
		\centering
		\includegraphics[width=1.05\textwidth]{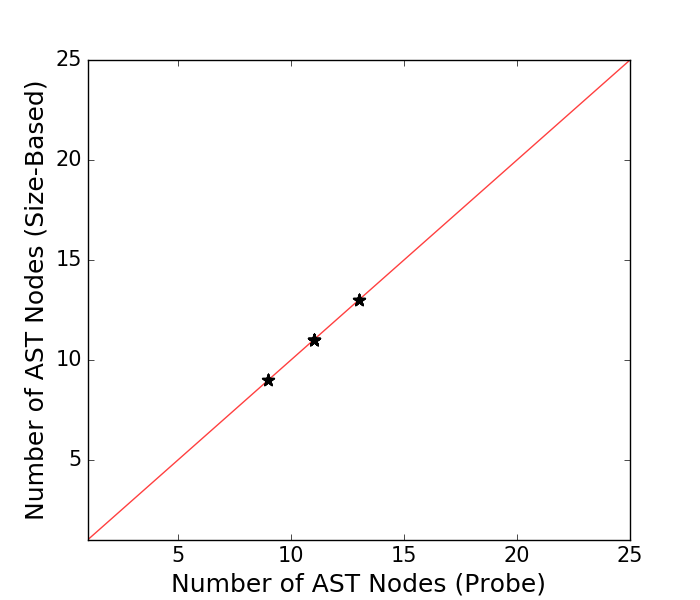}
		\caption{Size-based \circuit domain}
		\label{fig:sizebased-circuit}
	\end{subfigure}%
	\begin{subfigure}{.33\textwidth}
		\centering
		\includegraphics[width=1.05\textwidth]{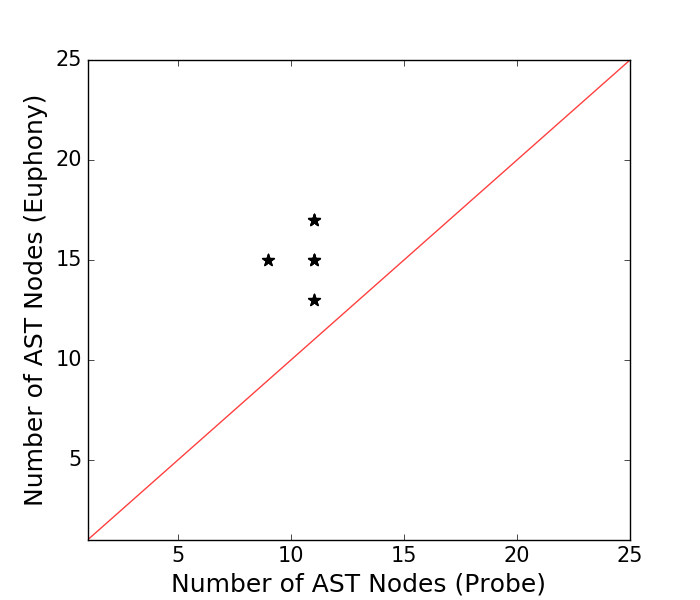}
		\caption{\euphony \circuit domain}
		\label{fig:euphony-circuit}
	\end{subfigure}%
	\begin{subfigure}{.33\textwidth}
		\centering
		\includegraphics[width=1.05\textwidth]{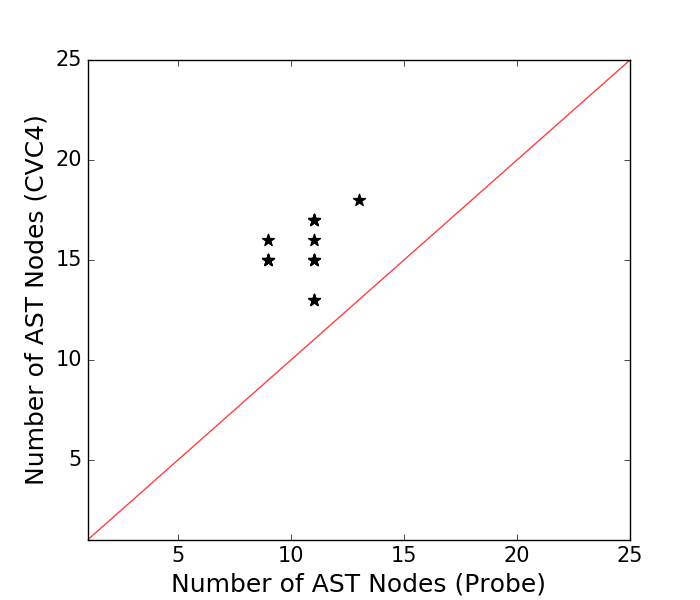}
		\caption{\cvc \circuit domain}
		\label{fig:cvc4-circuit}
	\end{subfigure}
	\caption{Comparison between sizes of programs generated by different algorithms. 
  \autoref{fig:sizebased-string}, \autoref{fig:euphony-string} and \autoref{fig:cvc4-string} compare \tool \vs size-based enumeration, \euphony and \cvc, respectively, 
  on the \stringbench domain; graphs are log scale. 
  \autoref{fig:sizebased-bitvec}, \autoref{fig:euphony-bitvec} and \autoref{fig:cvc4-bitvec} compare the same pairs of tools on the \bitvec domain 
  and \autoref{fig:sizebased-circuit}, \autoref{fig:euphony-circuit} and \autoref{fig:cvc4-circuit} on the \circuit domain; graphs are linear scale.}
	\label{fig:size-string}
\end{figure}

\mypara{Size of generated solutions}
\autoref{fig:size-string} shows the sizes of \tool solutions in AST nodes, 
as compared to size-based enumeration (which always returns the smallest solution by definition),
as well as \euphony and \cvc.
Each comparison is limited to the benchmarks both tools can solve.

\mypara{\stringbench domain}
First, we notice in \autoref{fig:sizebased-string} that \tool sometimes finds larger solutions than size-based enumeration, but the difference is small.
Likewise, \autoref{fig:euphony-string} shows that \euphony and \tool return similar-sized solutions. \tool returns the smaller solutions for 10 benchmarks, but the difference is not large.
On the other hand, \cvc solutions (\autoref{fig:cvc4-string}) are larger than \tool's on 41 out of 45 benchmarks, 
sometimes by as much as \emph{two orders of magnitude}. 
For the remaining four benchmarks, solution sizes are equal.
On one of the benchmarks not solved by \tool (and therefore not in the graph), \cvc returns a result with over 7100(!) AST nodes.

\mypara{Other domains}
\autoref{fig:sizebased-bitvec} shows that on the \bitvec domain \tool finds the minimal solution in all cases except one.
Solutions by \euphony (\autoref{fig:euphony-bitvec}) and \cvc (\autoref{fig:cvc4-bitvec}) 
are slightly larger\footnote{Note that we use linear scale for \bitvec and \circuit as opposed to logarithmic scale for \stringbench.} in one (resp. two) cases, but the difference is small.
For the \circuit benchmarks, \tool always finds minimal solutions, as shown in \autoref{fig:sizebased-circuit}. 
Both \euphony (\autoref{fig:euphony-circuit}) and \cvc (\autoref{fig:cvc4-circuit}) generate larger solutions for \textit{all} of the commonly solved benchmarks. 
Hence, on the \circuit domain, \tool outperforms its competitors with respect to \emph{both} synthesis time and solution size.

\begin{figure}[t]
	\centering
	\begin{subfigure}{.49\textwidth}
		\centering
		\includegraphics[width=.8\textwidth]{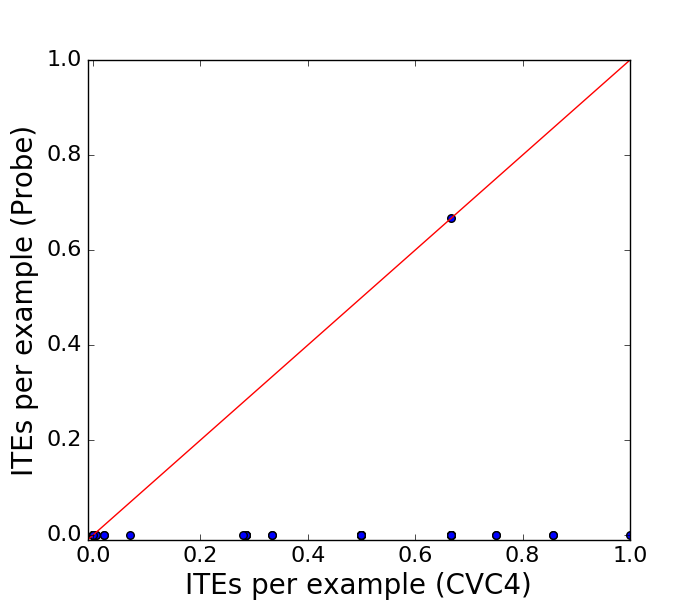}
		\caption{Number of \scode{ite} operations per examples}
		\label{fig:ite}
	\end{subfigure}%
	\begin{subfigure}{.48\textwidth}
		\centering
		\resizebox{\textwidth}{!}{
			\begin{tabular}{|c|c|c|c|c|}\hline
				& Training & Testing  & \tool    & \cvc      \\
				Test Benchmark & Examples & Examples & Accuracy & Accuracy  \\\hline
				initials-long & 4 & 54 & 100\% & 100\%            \\
				phone-5-long  & 7 & 100 & 100\% & 100\%            \\
				phone-6-long & 7 & 100 & 100\% & 100\%            \\
				phone-7-long & 7 & 100 & 100\% & 7\%              \\	
				phone-10-long & 7 & 100 & 100\% & 57\%         \\\hline
				phone-9-long & 7 & 100 & N/A & 7\%              \\
				univ\_4-long & 8 & 20& N/A &  73.6\%           \\
				univ\_5-long &  8 & 20& N/A & 68.4\%           \\
				univ\_6-long  & 8 & 20& N/A &  100\%            \\\hline	
				Avg Accuracy &&& 100\% & 68.1\%         \\\hline
			\end{tabular}
		}
		\caption{Generalization accuracy on unseen inputs}\label{tbl:accuracy}
	\end{subfigure}%
	\caption{\autoref{fig:ite} displays the number of \scode{ite} operations per example for the \stringbench benchmarks solved by \tool and \cvc. 
  \cvc has a large number of case splits as indicated. \autoref{tbl:accuracy} shows the generalization accuracy on unseen inputs for the 9 test benchmarks.}
	\label{fig:aux-string-expt}
\end{figure}

\mypara{Case splitting}
So why are the \cvc \stringbench programs so large?
Upon closer examination, we determined that they perform over-abundant \emph{case splitting},
which hurts both readability and generality.
To confirm our intuition, we count the number of if-then-else operations (\scode{ite})
in the programs synthesized by \tool and by \cvc. 
The results are plotted in \autoref{fig:ite}.
%
The number of \scode{ite}s is normalized by number of examples in the task specification.
\tool averages $0.01$ \scode{ite} per example (for all but one benchmark \tool solutions do not contain an \scode{ite}), 
whereas \cvc averages $0.42$ \scode{ite}s per example.
When also considering benchmarks \tool cannot solve, some \cvc programs have more than two \scode{ite}s per example.

\mypara{Generalization Accuracy}
Finally, we test the generality of the synthesized programs---%
whether they generalize well to additional examples, or in other words, whether synthesis returns reusable code.
Concretely, we measure \textit{generalization accuracy}~\cite{Alpaydin14}, the percentage of unseen inputs for which a generated program produces the correct output.
To this end, we find a solution using \tool and \cvc, and then test it on additional examples for the same program.

Since most benchmarks in our suite contain only a few input-output examples,
splitting these examples into a training and testing set would render most benchmarks severely underspecified.
Instead we turn to a subset of the \stringbench benchmarks from the \sygus Competition PBE-Strings suite.
These are benchmark pairs where each task appears in a ``short'' form with a small number of examples and a ``long'' form with additional examples,
but both represent the same task and share the same grammar.
There are nine such benchmark pairs in this suite.

We compare the generalization accuracy of \cvc and \tool by using the short benchmark of each pair to synthesize a solution,
and, if a solution is found, we test it on the examples of the long version of the benchmark to see how well it generalizes.
The results are shown in \autoref{tbl:accuracy}.

\begin{figure}[htp]
	\resizebox{\textwidth}{!}{
		\begin{tabular}{|c|c|c|}
			\hline
			\textbf{Benchmark}  & \textbf{Solution generated}  & \textbf{Time (s)}\\ \hline
			stackoverflow1.sl & \scode{(substr arg 0 (+ (indexof arg "Inc" 1) -1))} & 2.2s \\\hline
			stackoverflow3.sl & 
			\scode{(substr arg (- (to.int (concat "1" "9")) 2) (len arg))} & 2.1s\\\hline
			stackoverflow8.sl & \scode{(substr arg (- (len arg) (+ (+ 2 4) 4)) (len arg))} & 6.5s \\\hline
			stackoverflow10.sl & \scode{(substr arg (indexof (replace arg " " (to.str (len arg))) " " 1) 4)}&  27.6s\\\hline
			exceljet1.sl & \scode{(substr arg1 (+ (indexof arg1 "\_" 1) 1) (len arg1))}  & 1.5s \\\hline
			exceljet2.sl &
			\scode{(replace (substr arg (- (len arg) (indexof arg "." 1)) (len arg)) "." "")} & 16.5s\\\hline
			initials.sl &
			\scode{(concat (concat (at name 0) ".") (concat (at name (+ (indexof name " " 0) 1)) "."))} & 134.5s \\\hline
			phone-6-long.sl & 
			\scode{(substr name (- (indexof name "-" 4) 3) 3)} & 3.4s \\\hline
			43606446.sl & \scode{(substr arg (- (len arg) (+ (+ 1 1) (+ 1 1))) (+ (+ 1 1) 1))} & 10.8s \\\hline
			11604909.sl & \scode{(substr (concat " " arg) (indexof arg "." 1) (+ (+ 1 1) 1))}& 15.9s 
			\\\hline
		\end{tabular}
	}
	\caption{\tool solutions for 10 randomly selected benchmarks out of the 48 benchmarks \tool solves from the \cite{euphony} \stringbench testing set, Time indicates the synthesis time in seconds. }\label{tbl:sample-solution-string}
\end{figure}
\begin{figure}[htp]
	\resizebox{\textwidth}{!}{
		\begin{tabular}{|c|c|c|}
			\hline
			\textbf{Benchmark}  & \textbf{Solution generated}  & \textbf{Time (s)}\\ \hline
			hd-11.sl & \scode{(bvult y (bvand x (bvnot y)))} & 2.4s \\\hline
			hd-09.sl & 
			\scode{(bvsub x (bvshl (bvand (bvashr x \#x000000000000001f) x) \#x0000000000000001))} &  6.3s\\\hline
			hd-15.sl & \scode{(bvsub (bvor x y) (bvlshr (bvxor x y) \#x0000000000000001))} & 13s \\\hline
			hd-18.sl & \scode{(bvult (bvxor x (bvneg x)) (bvneg x))}&  1.3s\\\hline
			hd-13.sl & \scode{(bvor (bvashr x \#x000000000000001f) (bvlshr (bvneg x) \#x000000000000001f))}  & 1.6s \\\hline
		\end{tabular}
	}
	\caption{\tool solutions for 5 randomly selected benchmarks out of the 21 benchmarks \tool solves from the Hacker's Delight \bitvec set, Time indicates the synthesis time in seconds. }\label{tbl:sample-solution-bitvec}
\end{figure}

\begin{figure}[htp]
	\resizebox{\textwidth}{!}{
		\begin{tabular}{|c|c|c|}
			\hline
			\textbf{Benchmark}  & \textbf{Solution generated}  & \textbf{Time (s)}\\ \hline
			CrCy\_10-sbox2-D5-sIn14.sl & \scode{(xor LN200 (xor LN61 (and (xor LN16 LN17) LN4)))} & 9.4s \\\hline
			CrCy\_10-sbox2-D5-sIn88.sl & 
			\scode{(xor LN73 (and (xor (and LN70 (xor (xor LN236 LN252) LN253)) LN71) LN74))} &  287.1s\\\hline
			CrCy\_10-sbox2-D5-sIn78.sl & \scode{(and (xor (and LN70 (xor (xor LN236 LN252) LN253)) LN73) LN77)} & 11.8s \\\hline
			CrCy\_10-sbox2-D5-sIn80.sl & \scode{(xor LN73 (and LN70 (xor (xor LN236 LN252) LN253)))}&  2.2s\\\hline
			CrCy\_8-P12-D5-sIn1.sl & \scode{(xor (xor (xor LN3 LN7) (xor (xor LN75 LN78) LN81)) k4)}  & 9.1s \\\hline
		\end{tabular}
	}
	\caption{\tool solutions for 5 randomly selected benchmarks out of the 22 benchmarks \tool solves from the \cite{euphony} \circuit set, Time indicates the synthesis time in seconds. }\label{tbl:sample-solution-circuit}
\end{figure}

The first part of the table shows the benchmarks where \tool finds a solution. 
As discussed above, \tool rarely finds solutions with case splits, so it is not surprising that once it finds a program, 
that program is not at all overfitted to the examples.

Solutions found by \cvc generalize with 100\% accuracy in 4 out of the 9 benchmark pairs. 
In two of the benchmarks, the accuracy of \cvc solutions is only 7\%, 
or precisely the 7 training examples out of the 100-example test set, 
representing a complete overfitting to the training examples.
On average, \cvc has 68\% generalization accuracy on these benchmark pairs.
Even though this experiment is small, 
it provides a glimpse into the extent to which \cvc solutions sometimes overfit to the examples.

\mypara{Sample solutions}
Finally, we examine a few sample solutions generated by \tool in \autoref{tbl:sample-solution-string} for the \stringbench domain, \autoref{tbl:sample-solution-bitvec} for the \bitvec domain and \autoref{tbl:sample-solution-circuit} for the \circuit domain.
Even though the \sygus language is unfamiliar to most readers, 
we believe that these solutions should appear simple and clearly understandable. 
In comparison, the \cvc solutions to these benchmarks are dozens or hundreds of operations long.

\mypara{Solution quality}
The experiments in this section explored solution quality via three empirical measures: solution size, 
the number of case-splits, 
and the ability of solutions to generalize to new examples for the same task.
These results show conclusively that, while \cvc is considerably faster than \tool, and solves more benchmarks, 
the quality of its solutions is significantly worse.

\subsection{Conclusions}

In conclusion, we have shown that \tool is faster and solves more benchmarks than 
unguided enumerative techniques, 
which confirms that just-in-time learning is an improvement on a baseline synthesizer. 
We have also shown that \tool is faster and solves more benchmarks than \euphony, 
a probabilistic synthesizer with a pre-learned model,
based on top-down enumeration.
Finally, we have explored the quality of synthesized solutions via size, case splitting, and generalizability, 
and found that even though \cvc solves more benchmarks than \tool, 
its solutions to example-based benchmarks overfit to the examples, 
and are therefore neither readable nor reusable; 
in contrast, \tool's solutions are small and generalize perfectly.
\section{Related Work}\label{sec:related}

\mypara{Enumerative Program Synthesis}
Despite their simplicity, enumerative program synthesizers are known to be very effective:
\esolver~\cite{sygus} and \eusolver~\cite{alur2017scaling} have been past winners of the \sygus competition~\cite{alur2017sygus, alur2016sygus}.
Enumerative synthesizers typically explore the space of programs either top-down,
by extending a partial program tree from the node towards the leaves~\cite{lee2018accelerating,alur2017scaling,kalyan2018neural,koukoutos2017repair}, 
or bottom-up, by gradually building up a program tree from the leaves towards the root~\cite{udupa2013transit,albarghouthi2013recursive,sygus,peleg2020perfect}. 
%
These two strategies have complementary strengths and weaknesses,
similar to backward chaining and forward chaining in proof search.
%

One important advantage of bottom-up enumeration for inductive synthesis
is the ability to prune the search space using \emph{observational equivalence} (OE),
\ie discard a program that behaves equivalently to an already enumerated program 
on the set of inputs from the semantic specification.
OE was first proposed in~\cite{udupa2013transit,albarghouthi2013recursive}
and since then has been successfully used in many bottom-up synthesizers~\cite{wang2017synthesizing,peleg2020perfect,cacm2018searchbased},
including \tool.
Top-down enumeration techniques cannot fully leverage OE,
because incomplete programs they generate cannot be evaluated on the inputs.
Instead, these synthesizers prune the space based on other syntactic and semantic notions of program equivalence:
for example,
\cite{gvero2013complete,osera2015type,Frankle2016ExampleDirected} only produce programs in a normal form;
\cite{feser2015synthesizing,Kneuss2013synthesis,SmithA19Equivalence} perform symmetry reduction based on equational theories (either built-in or user-provided);
finally, \euphony~\cite{lee2018accelerating} employs a weaker version of OE for incomplete programs,
which compares their complete parts observationally and their incomplete parts syntactically.

\mypara{Guiding Synthesis with Probabilistic Models}
Recent years have seen proliferation of probabilistic models of programs~\cite{allamanis2018survey}, 
which can be used, in particular, to guide program synthesis. 
The general idea is to prioritize the exploration of grammar productions
based on scores assigned by a probabilistic model;
the specific technique, however, varies depending on 
\begin{inparaenum}[(1)]
\item the context taken into consideration by the model when assigning scores, and
\item how the scores are taken into account during search. 
\end{inparaenum}
Like \tool, \cite{koukoutos2017repair, balog2016deepcoder, menon2013machine} use a PCFG,
which assigns scores to productions \emph{independently of their context} within the synthesized program;
unlike \tool, however, these techniques select the PCFG once, at the beginning of the synthesis process, 
based on a learned mapping from semantic specifications to scores.
On the opposite end of the spectrum, \metal~\cite{si2018learning} and \concord~\cite{chen2020program} 
use graph-based and sequence-based models, respectively, to condition the scores
on the \emph{entire partial program} that is being extended.
In between these extremes,
\euphony uses a learned context in the form of a \emph{probabilistic higher-order grammar}~\cite{bielik2016phog},
while NGDS~\cite{kalyan2018neural} conditions the scores on the \emph{local specification} propagated top-down by the deductive synthesizer.
The more context a model takes into account, 
the more precise the guidance it provides, but also the harder it is to learn.
Another consideration is that neural models, used in~\cite{kalyan2018neural,si2018learning,chen2020program}
incur a larger overhead than simple grammar-based models, used in \tool and~\cite{menon2013machine,balog2016deepcoder,koukoutos2017repair,lee2018accelerating},
since they have to invoke a neural network at each branching point during search. 

As for using the scores to guide search,
most existing techniques are specific to \emph{top-down enumeration}.
They include prioritized depth-first search~\cite{balog2016deepcoder},
branch and bound search~\cite{kalyan2018neural},
and variants of best-first search~\cite{menon2013machine,lee2018accelerating,koukoutos2017repair}.
In contrast to these approaches, \tool uses the scores to guide \emph{bottom-up enumeration} with observational equivalence reduction.
\tool's enumeration is essentially a bottom-up version of best-first search,
and it empirically performs better than the top-down best-first search in \euphony;
one limitation, however, is that our algorithm is specific to PCFGs 
and extending it to models that require more context is not straightforward.

\deepcoder~\cite{balog2016deepcoder} also proposes a scheme they call \emph{sort and add},
which is not specific to top-down enumeration and can be used in conjunction with any synthesis algorithm:
this scheme runs synthesis with a reduced grammar, containing only productions with highest scores,
and iteratively adds less likely productions if no solution is found.
Although very general, this scheme is less efficient than best-first search:
it can waste resources searching with an insufficient grammar,
and has to revisit the same programs again once the search is restarted with a larger grammar.


Finally, \metal and \concord, which are based on reinforcement learning (RL), 
do not perform traditional backtracking search at all.
Instead, at each branching point, they simply choose a single production that has the highest score according to the current RL policy;
a sequence of such decisions is called a \emph{policy rollout}.
If a rollout does not lead to a solution,
the policy is updated according to a reward function explained below and a new rollout is performed from scratch.

\mypara{Learning Probabilistic Models}
Approaches to \emph{learning} probabilistic models of programs can be classified into two categories:
pre-training and learning on the fly.
In the first category, \cite{menon2013machine}, \euphony, and \ngds
are trained using a large corpus of human-designed synthesis problems and their gold standard solutions
(the latter can be provided by a human or synthesized using size-based enumeration).
Such datasets are costly to obtain:
because these models are domain-specific, a new training corpus has to be designed for each domain.
In contrast, \deepcoder learns from randomly sampled programs and inputs;
it is, however, unclear how effective this technique is for domains beyond the highly restricted DSL in the paper.
Unlike all these approaches, \tool requires no pre-training,
and hence can be used on a new domain without any up-front cost;
if a pre-trained PCFG for the domain is available, however, 
\tool can also be initialized with this model (although we have not explored this avenue in the present work).


\dreamcoder, \metal, and \concord are related to the just-in-time approach of \tool in the sense that they update their probabilistic model on the fly. 
\dreamcoder learns a probabilistic model from \emph{full solutions} to a subset of synthesis problems from a corpus, 
whereas \tool learns a problem-specific model from \emph{partial solutions} to a single synthesis problem.

The RL-based tools \metal and \concord start with a pre-trained RL policy
and then fine-tune it for the specific task during synthesis.
Note that off-line training is vital for the performance of these tools,
while \tool is effective even without a pre-trained model.
The \emph{reward mechanism} in \metal is similar to \tool: 
it rewards a policy based on the fraction of input-output examples solved by its rollout.
\concord instead rewards its policies based on infeasibility information from a deductive reasoning engine:
productions that expand to infeasible programs have lower probability in the next rollout.
Although the \concord paper reports that its reward mechanism outperforms that of \metal,
we conjecture that rewards based on partial solutions are simply not as good a fit for RL
as they are for bottom-up enumeration:
as we discuss in \autoref{sec:partial}, it is crucial to learn from \emph{shortest} partial solutions
to avoid irrelevant syntactic features;
policy rollouts do not guarantee that short solutions are generated first.
Finally, \concord's reward mechanism requires expensive solver invocations to check infeasibility of partial programs,
while \tool's reward mechanism incurs practically no overhead compared to unguided search.

\mypara{Leveraging Partial Solutions to Guide Synthesis}
\lasy~\cite{perelman2014test} and \frangel~\cite{shi2019frangel} are component-based synthesis techniques that leverage information from partial solutions to generate new programs. 
\lasy explicitly requires the user to arrange input-output examples in the order of increasing difficulty,
and then synthesizes a sequence of programs, where $i$\textsuperscript{th} program passes the first $i$ examples.
Each following program is not synthesized from scratch,
but rather by modifying the previous program;
hence intermediate programs serve as ``stepping stones'' for synthesis.
\tool puts less burden on the user:
it does not require the examples to be arranged in a sequence,
and instead identifies partial solutions that satisfy any subset of examples. 

 
Similar to \tool, \frangel leverages partial solutions that satisfy any subset of the example specification. 
\frangel generates new programs by randomly combining fragments from partial solutions.
\tool is similar to \frangel and \lasy in that it guides the search using syntactic information learned from partial solutions, 
but we achieve that by updating the weights of useful productions in a probabilistic grammar and using it to guide bottom-up enumerative search.

Our previous work, \bester~\cite{peleg2020perfect} proposes a technique to accumulate multiple partial solutions during bottom-up enumerative synthesis with minimum overhead. \tool is a natural extension of \bester: it leverages these accumulated partial solutions to guide search.

During top-down enumeration, \cite{koukoutos2017repair} employs an optimization strategy where the cost of an incomplete (partial) program is lowered if it satisfies some of the examples. This optimization encourages the search to complete a partial program that looks promising, but unlike \tool, offers no guidance on which are the likely productions to complete it with. Moreover, this optimization only works on partial programs that can be evaluated on some examples. \tool's bottom-up search generates complete programs that can always be evaluated on all examples.


\section{Conclusion and Future work}

We have presented a new program synthesis algorithm we dub \emph{guided bottom-up search with just-in-time-learning}.
This algorithm combines the pruning power of observational equivalence 
with guidance from probabilistic models.
Moreover, our just-in-time learning is able to bootstrap a probabilistic model during synthesis by leveraging partial solutions,
and hence does not require training data, which can be hard to obtain.

We have implemented this algorithm in a tool called \tool 
that works with the popular \sygus input format. 
We evaluated \tool on 140 synthesis benchmarks from three different domains. 
Our evaluation demonstrates that \tool is more efficient than unguided enumerative search 
and a state-of-the-art guided synthesizer \euphony,
and while \tool is less efficient than \cvc, our solutions are of higher quality.

In future work, we are interested in instantiating \tool in new application domains. 
We expect just-in-time learning to work for programs over structured data structures,
\eg lists and tree transformations. 
Just-in-time learning also requires that example specifications cover a range from simple to more complex, 
so that \tool can discover short partial solutions and learn from them. 
Luckily, users seem to naturally provide examples that satisfy this property, 
as indicated by \sygus benchmarks whose specifications are taken from StackOverflow. 
Generalizing these observations is an exciting direction for future work.
Another interesting direction is to consider \tool in the context of program repair, 
where similarity to the original faulty program can serve as a prior to initialize the PCFG.

\begin{acks}
	The authors would like to thank the anonymous reviewers for their feedback on the draft of this paper.
	This work was supported by the National Science Foundation under Grants No.~1955457, 1911149, and 1943623.
\end{acks}  
	
	\bibliography{library}


\begin{thebibliography}{48}


\ifx \showCODEN    \undefined \def \showCODEN     #1{\unskip}     \fi
\ifx \showDOI      \undefined \def \showDOI       #1{#1}\fi
\ifx \showISBNx    \undefined \def \showISBNx     #1{\unskip}     \fi
\ifx \showISBNxiii \undefined \def \showISBNxiii  #1{\unskip}     \fi
\ifx \showISSN     \undefined \def \showISSN      #1{\unskip}     \fi
\ifx \showLCCN     \undefined \def \showLCCN      #1{\unskip}     \fi
\ifx \shownote     \undefined \def \shownote      #1{#1}          \fi
\ifx \showarticletitle \undefined \def \showarticletitle #1{#1}   \fi
\ifx \showURL      \undefined \def \showURL       {\relax}        \fi
\providecommand\bibfield[2]{#2}
\providecommand\bibinfo[2]{#2}
\providecommand\natexlab[1]{#1}
\providecommand\showeprint[2][]{arXiv:#2}

\bibitem[\protect\citeauthoryear{??}{eup}{2018}]%
        {euphony}
 \bibinfo{year}{2018}\natexlab{}.
\newblock \bibinfo{title}{Euphony Benchmark Suite}.
\newblock
\newblock
\urldef\tempurl%
\url{https://github.com/wslee/euphony/tree/master/benchmarks}
\showURL{%
\tempurl}


\bibitem[\protect\citeauthoryear{Albarghouthi, Gulwani, and
  Kincaid}{Albarghouthi et~al\mbox{.}}{2013}]%
        {albarghouthi2013recursive}
\bibfield{author}{\bibinfo{person}{Aws Albarghouthi}, \bibinfo{person}{Sumit
  Gulwani}, {and} \bibinfo{person}{Zachary Kincaid}.}
  \bibinfo{year}{2013}\natexlab{}.
\newblock \showarticletitle{Recursive program synthesis}. In
  \bibinfo{booktitle}{\emph{International Conference on Computer Aided
  Verification}}. Springer, \bibinfo{pages}{934--950}.
\newblock


\bibitem[\protect\citeauthoryear{Allamanis, Barr, Devanbu, and
  Sutton}{Allamanis et~al\mbox{.}}{2018}]%
        {allamanis2018survey}
\bibfield{author}{\bibinfo{person}{Miltiadis Allamanis},
  \bibinfo{person}{Earl~T Barr}, \bibinfo{person}{Premkumar Devanbu}, {and}
  \bibinfo{person}{Charles Sutton}.} \bibinfo{year}{2018}\natexlab{}.
\newblock \showarticletitle{A survey of machine learning for big code and
  naturalness}.
\newblock \bibinfo{journal}{\emph{ACM Computing Surveys (CSUR)}}
  \bibinfo{volume}{51}, \bibinfo{number}{4} (\bibinfo{year}{2018}),
  \bibinfo{pages}{1--37}.
\newblock


\bibitem[\protect\citeauthoryear{Alpaydin}{Alpaydin}{2014}]%
        {Alpaydin14}
\bibfield{author}{\bibinfo{person}{Ethem Alpaydin}.}
  \bibinfo{year}{2014}\natexlab{}.
\newblock \bibinfo{booktitle}{\emph{Introduction to Machine Learning}
  (\bibinfo{edition}{3} ed.)}.
\newblock \bibinfo{publisher}{MIT Press}, \bibinfo{address}{Cambridge, MA}.
\newblock
\showISBNx{978-0-262-02818-9}


\bibitem[\protect\citeauthoryear{Alur, Bod{\'{\i}}k, Juniwal, Martin,
  Raghothaman, Seshia, Singh, Solar{-}Lezama, Torlak, and Udupa}{Alur
  et~al\mbox{.}}{2013}]%
        {sygus}
\bibfield{author}{\bibinfo{person}{Rajeev Alur}, \bibinfo{person}{Rastislav
  Bod{\'{\i}}k}, \bibinfo{person}{Garvit Juniwal}, \bibinfo{person}{Milo M.~K.
  Martin}, \bibinfo{person}{Mukund Raghothaman}, \bibinfo{person}{Sanjit~A.
  Seshia}, \bibinfo{person}{Rishabh Singh}, \bibinfo{person}{Armando
  Solar{-}Lezama}, \bibinfo{person}{Emina Torlak}, {and}
  \bibinfo{person}{Abhishek Udupa}.} \bibinfo{year}{2013}\natexlab{}.
\newblock \showarticletitle{Syntax-guided synthesis}. In
  \bibinfo{booktitle}{\emph{Formal Methods in Computer-Aided Design, {FMCAD}
  2013, Portland, OR, USA, October 20-23, 2013}}. \bibinfo{pages}{1--8}.
\newblock
\urldef\tempurl%
\url{http://ieeexplore.ieee.org/document/6679385/}
\showURL{%
\tempurl}


\bibitem[\protect\citeauthoryear{Alur, Fisman, Singh, and Solar-Lezama}{Alur
  et~al\mbox{.}}{2016}]%
        {alur2016sygus}
\bibfield{author}{\bibinfo{person}{Rajeev Alur}, \bibinfo{person}{Dana Fisman},
  \bibinfo{person}{Rishabh Singh}, {and} \bibinfo{person}{Armando
  Solar-Lezama}.} \bibinfo{year}{2016}\natexlab{}.
\newblock \showarticletitle{Sygus-comp 2016: results and analysis}.
\newblock \bibinfo{journal}{\emph{arXiv preprint arXiv:1611.07627}}
  (\bibinfo{year}{2016}).
\newblock


\bibitem[\protect\citeauthoryear{Alur, Fisman, Singh, and Solar-Lezama}{Alur
  et~al\mbox{.}}{2017a}]%
        {alur2017sygus}
\bibfield{author}{\bibinfo{person}{Rajeev Alur}, \bibinfo{person}{Dana Fisman},
  \bibinfo{person}{Rishabh Singh}, {and} \bibinfo{person}{Armando
  Solar-Lezama}.} \bibinfo{year}{2017}\natexlab{a}.
\newblock \showarticletitle{Sygus-comp 2017: Results and analysis}.
\newblock \bibinfo{journal}{\emph{arXiv preprint arXiv:1711.11438}}
  (\bibinfo{year}{2017}).
\newblock


\bibitem[\protect\citeauthoryear{Alur, Radhakrishna, and Udupa}{Alur
  et~al\mbox{.}}{2017b}]%
        {alur2017scaling}
\bibfield{author}{\bibinfo{person}{Rajeev Alur}, \bibinfo{person}{Arjun
  Radhakrishna}, {and} \bibinfo{person}{Abhishek Udupa}.}
  \bibinfo{year}{2017}\natexlab{b}.
\newblock \showarticletitle{Scaling enumerative program synthesis via divide
  and conquer}. In \bibinfo{booktitle}{\emph{{International Conference on Tools
  and Algorithms for the Construction and Analysis of Systems}}}. Springer,
  \bibinfo{pages}{319--336}.
\newblock


\bibitem[\protect\citeauthoryear{Alur, Singh, Fisman, and Solar-Lezama}{Alur
  et~al\mbox{.}}{2018}]%
        {cacm2018searchbased}
\bibfield{author}{\bibinfo{person}{Rajeev Alur}, \bibinfo{person}{Rishabh
  Singh}, \bibinfo{person}{Dana Fisman}, {and} \bibinfo{person}{Armando
  Solar-Lezama}.} \bibinfo{year}{2018}\natexlab{}.
\newblock \showarticletitle{Search-based Program Synthesis}.
\newblock \bibinfo{journal}{\emph{Commun. ACM}} \bibinfo{volume}{61},
  \bibinfo{number}{12} (\bibinfo{date}{Nov.} \bibinfo{year}{2018}),
  \bibinfo{pages}{84--93}.
\newblock
\showISSN{0001-0782}
\urldef\tempurl%
\url{https://doi.org/10.1145/3208071}
\showDOI{\tempurl}


\bibitem[\protect\citeauthoryear{Balog, Gaunt, Brockschmidt, Nowozin, and
  Tarlow}{Balog et~al\mbox{.}}{2016}]%
        {balog2016deepcoder}
\bibfield{author}{\bibinfo{person}{Matej Balog}, \bibinfo{person}{Alexander~L
  Gaunt}, \bibinfo{person}{Marc Brockschmidt}, \bibinfo{person}{Sebastian
  Nowozin}, {and} \bibinfo{person}{Daniel Tarlow}.}
  \bibinfo{year}{2016}\natexlab{}.
\newblock \showarticletitle{Deepcoder: Learning to write programs}.
\newblock \bibinfo{journal}{\emph{arXiv preprint arXiv:1611.01989}}
  (\bibinfo{year}{2016}).
\newblock


\bibitem[\protect\citeauthoryear{Barke, Peleg, and Polikarpova}{Barke
  et~al\mbox{.}}{2020}]%
        {extended}
\bibfield{author}{\bibinfo{person}{Shraddha Barke}, \bibinfo{person}{Hila
  Peleg}, {and} \bibinfo{person}{Nadia Polikarpova}.}
  \bibinfo{year}{2020}\natexlab{}.
\newblock \showarticletitle{Just-in-Time Learning for Bottom-up Enumerative
  Synthesis}.
\newblock  (\bibinfo{year}{2020}).
\newblock
\urldef\tempurl%
\url{https://shraddhabarke.github.io/publication/probe-oopsla}
\showURL{%
\tempurl}


\bibitem[\protect\citeauthoryear{Bielik, Raychev, and Vechev}{Bielik
  et~al\mbox{.}}{2016}]%
        {bielik2016phog}
\bibfield{author}{\bibinfo{person}{Pavol Bielik}, \bibinfo{person}{Veselin
  Raychev}, {and} \bibinfo{person}{Martin Vechev}.}
  \bibinfo{year}{2016}\natexlab{}.
\newblock \showarticletitle{PHOG: probabilistic model for code}. In
  \bibinfo{booktitle}{\emph{International Conference on Machine Learning}}.
  \bibinfo{pages}{2933--2942}.
\newblock


\bibitem[\protect\citeauthoryear{Chen, Wang, Bastani, Dillig, and Feng}{Chen
  et~al\mbox{.}}{2020}]%
        {chen2020program}
\bibfield{author}{\bibinfo{person}{Yanju Chen}, \bibinfo{person}{Chenglong
  Wang}, \bibinfo{person}{Osbert Bastani}, \bibinfo{person}{Isil Dillig}, {and}
  \bibinfo{person}{Yu Feng}.} \bibinfo{year}{2020}\natexlab{}.
\newblock \showarticletitle{Program Synthesis Using Deduction-Guided
  Reinforcement Learning}. In \bibinfo{booktitle}{\emph{International
  Conference on Computer Aided Verification}}. Springer,
  \bibinfo{pages}{587--610}.
\newblock


\bibitem[\protect\citeauthoryear{Ellis, Morales, Meyer, Solar-Lezama, and
  Tenenbaum}{Ellis et~al\mbox{.}}{2018}]%
        {ellis2018search}
\bibfield{author}{\bibinfo{person}{Kevin Ellis}, \bibinfo{person}{Lucas
  Morales}, \bibinfo{person}{Mathias~Sabl{\'e} Meyer}, \bibinfo{person}{Armando
  Solar-Lezama}, {and} \bibinfo{person}{Joshua~B Tenenbaum}.}
  \bibinfo{year}{2018}\natexlab{}.
\newblock \showarticletitle{Search, compress, compile: Library learning in
  neurally-guided bayesian program learning}.
\newblock \bibinfo{journal}{\emph{Advances in neural information processing
  systems}} (\bibinfo{year}{2018}).
\newblock


\bibitem[\protect\citeauthoryear{Feng, Martins, Geffen, Dillig, and
  Chaudhuri}{Feng et~al\mbox{.}}{2017a}]%
        {FengMGDC17}
\bibfield{author}{\bibinfo{person}{Yu Feng}, \bibinfo{person}{Ruben Martins},
  \bibinfo{person}{Jacob~Van Geffen}, \bibinfo{person}{Isil Dillig}, {and}
  \bibinfo{person}{Swarat Chaudhuri}.} \bibinfo{year}{2017}\natexlab{a}.
\newblock \showarticletitle{Component-based synthesis of table consolidation
  and transformation tasks from examples}. In
  \bibinfo{booktitle}{\emph{Proceedings of the 38th {ACM} {SIGPLAN} Conference
  on Programming Language Design and Implementation, {PLDI} 2017, Barcelona,
  Spain, June 18-23, 2017}}. \bibinfo{pages}{422--436}.
\newblock


\bibitem[\protect\citeauthoryear{Feng, Martins, Wang, Dillig, and Reps}{Feng
  et~al\mbox{.}}{2017b}]%
        {petrinetsynth17}
\bibfield{author}{\bibinfo{person}{Yu Feng}, \bibinfo{person}{Ruben Martins},
  \bibinfo{person}{Yuepeng Wang}, \bibinfo{person}{Isil Dillig}, {and}
  \bibinfo{person}{Thomas~W Reps}.} \bibinfo{year}{2017}\natexlab{b}.
\newblock \showarticletitle{Component-based synthesis for complex APIs}.
\newblock \bibinfo{journal}{\emph{ACM SIGPLAN Notices}} \bibinfo{volume}{52},
  \bibinfo{number}{1} (\bibinfo{year}{2017}), \bibinfo{pages}{599--612}.
\newblock


\bibitem[\protect\citeauthoryear{Feser, Chaudhuri, and Dillig}{Feser
  et~al\mbox{.}}{2015}]%
        {feser2015synthesizing}
\bibfield{author}{\bibinfo{person}{John~K Feser}, \bibinfo{person}{Swarat
  Chaudhuri}, {and} \bibinfo{person}{Isil Dillig}.}
  \bibinfo{year}{2015}\natexlab{}.
\newblock \showarticletitle{Synthesizing data structure transformations from
  input-output examples}. In \bibinfo{booktitle}{\emph{ACM SIGPLAN Notices}},
  Vol.~\bibinfo{volume}{50}. ACM, \bibinfo{pages}{229--239}.
\newblock


\bibitem[\protect\citeauthoryear{Frankle, Osera, Walker, and Zdancewic}{Frankle
  et~al\mbox{.}}{2016}]%
        {Frankle2016ExampleDirected}
\bibfield{author}{\bibinfo{person}{Jonathan Frankle},
  \bibinfo{person}{Peter-Michael Osera}, \bibinfo{person}{David Walker}, {and}
  \bibinfo{person}{Steve Zdancewic}.} \bibinfo{year}{2016}\natexlab{}.
\newblock \showarticletitle{Example-directed Synthesis: A Type-theoretic
  Interpretation}. In \bibinfo{booktitle}{\emph{Proceedings of the 43rd Annual
  ACM SIGPLAN-SIGACT Symposium on Principles of Programming Languages}} (St.
  Petersburg, FL, USA) \emph{(\bibinfo{series}{POPL '16})}.
  \bibinfo{publisher}{ACM}, \bibinfo{address}{New York, NY, USA},
  \bibinfo{pages}{802--815}.
\newblock
\showISBNx{978-1-4503-3549-2}
\urldef\tempurl%
\url{https://doi.org/10.1145/2837614.2837629}
\showDOI{\tempurl}


\bibitem[\protect\citeauthoryear{Gao, Zhao, Ren, Swami, Ramanathan, and
  Bar{-}Noy}{Gao et~al\mbox{.}}{2012}]%
        {hypergraphs}
\bibfield{author}{\bibinfo{person}{Jianhang Gao}, \bibinfo{person}{Qing Zhao},
  \bibinfo{person}{Wei Ren}, \bibinfo{person}{Ananthram Swami},
  \bibinfo{person}{Ram Ramanathan}, {and} \bibinfo{person}{Amotz Bar{-}Noy}.}
  \bibinfo{year}{2012}\natexlab{}.
\newblock \showarticletitle{Dynamic Shortest Path Algorithms for Hypergraphs}.
\newblock \bibinfo{journal}{\emph{CoRR}}  \bibinfo{volume}{abs/1202.0082}
  (\bibinfo{year}{2012}).
\newblock
\showeprint[arxiv]{1202.0082}
\urldef\tempurl%
\url{http://arxiv.org/abs/1202.0082}
\showURL{%
\tempurl}


\bibitem[\protect\citeauthoryear{Gulwani}{Gulwani}{2011}]%
        {Gulwani:2011:ASP:1926385.1926423}
\bibfield{author}{\bibinfo{person}{Sumit Gulwani}.}
  \bibinfo{year}{2011}\natexlab{}.
\newblock \showarticletitle{Automating String Processing in Spreadsheets Using
  Input-output Examples}. In \bibinfo{booktitle}{\emph{Proceedings of the 38th
  Annual ACM SIGPLAN-SIGACT Symposium on Principles of Programming Languages}}
  (Austin, Texas, USA) \emph{(\bibinfo{series}{POPL '11})}.
  \bibinfo{publisher}{ACM}, \bibinfo{address}{New York, NY, USA},
  \bibinfo{pages}{317--330}.
\newblock
\showISBNx{978-1-4503-0490-0}
\urldef\tempurl%
\url{https://doi.org/10.1145/1926385.1926423}
\showDOI{\tempurl}


\bibitem[\protect\citeauthoryear{Gulwani}{Gulwani}{2016}]%
        {gulwani2016pbe}
\bibfield{author}{\bibinfo{person}{Sumit Gulwani}.}
  \bibinfo{year}{2016}\natexlab{}.
\newblock \showarticletitle{Programming by Examples (and its applications in
  Data Wrangling)}.
\newblock In \bibinfo{booktitle}{\emph{Verification and Synthesis of Correct
  and Secure Systems}}, \bibfield{editor}{\bibinfo{person}{Javier Esparza},
  \bibinfo{person}{Orna Grumberg}, {and} \bibinfo{person}{Salomon Sickert}}
  (Eds.). \bibinfo{publisher}{{IOS} Press}.
\newblock


\bibitem[\protect\citeauthoryear{Gulwani, Jha, Tiwari, and Venkatesan}{Gulwani
  et~al\mbox{.}}{2011}]%
        {gulwani2011synthesis}
\bibfield{author}{\bibinfo{person}{Sumit Gulwani}, \bibinfo{person}{Susmit
  Jha}, \bibinfo{person}{Ashish Tiwari}, {and} \bibinfo{person}{Ramarathnam
  Venkatesan}.} \bibinfo{year}{2011}\natexlab{}.
\newblock \showarticletitle{Synthesis of loop-free programs}.
\newblock \bibinfo{journal}{\emph{ACM SIGPLAN Notices}} \bibinfo{volume}{46},
  \bibinfo{number}{6} (\bibinfo{year}{2011}), \bibinfo{pages}{62--73}.
\newblock


\bibitem[\protect\citeauthoryear{Gvero, Kuncak, Kuraj, and Piskac}{Gvero
  et~al\mbox{.}}{2013}]%
        {gvero2013complete}
\bibfield{author}{\bibinfo{person}{Tihomir Gvero}, \bibinfo{person}{Viktor
  Kuncak}, \bibinfo{person}{Ivan Kuraj}, {and} \bibinfo{person}{Ruzica
  Piskac}.} \bibinfo{year}{2013}\natexlab{}.
\newblock \showarticletitle{Complete completion using types and weights}. In
  \bibinfo{booktitle}{\emph{ACM SIGPLAN Notices}}, Vol.~\bibinfo{volume}{48}.
  ACM, \bibinfo{pages}{27--38}.
\newblock


\bibitem[\protect\citeauthoryear{Inala and Singh}{Inala and Singh}{2018}]%
        {InalaS18}
\bibfield{author}{\bibinfo{person}{Jeevana~Priya Inala} {and}
  \bibinfo{person}{Rishabh Singh}.} \bibinfo{year}{2018}\natexlab{}.
\newblock \showarticletitle{{WebRelate:\ integrating web data with spreadsheets
  using examples}}.
\newblock \bibinfo{journal}{\emph{{PACMPL}}} \bibinfo{volume}{2},
  \bibinfo{number}{{POPL}} (\bibinfo{year}{2018}), \bibinfo{pages}{2:1--2:28}.
\newblock
\urldef\tempurl%
\url{{https://dl.acm.org/doi/10.1145/3158090}}
\showURL{%
\tempurl}


\bibitem[\protect\citeauthoryear{Jha, Gulwani, Seshia, and Tiwari}{Jha
  et~al\mbox{.}}{2010}]%
        {jha2010oracle}
\bibfield{author}{\bibinfo{person}{Susmit Jha}, \bibinfo{person}{Sumit
  Gulwani}, \bibinfo{person}{Sanjit~A Seshia}, {and} \bibinfo{person}{Ashish
  Tiwari}.} \bibinfo{year}{2010}\natexlab{}.
\newblock \showarticletitle{Oracle-guided component-based program synthesis}.
  In \bibinfo{booktitle}{\emph{2010 ACM/IEEE 32nd International Conference on
  Software Engineering}}, Vol.~\bibinfo{volume}{1}. IEEE,
  \bibinfo{pages}{215--224}.
\newblock


\bibitem[\protect\citeauthoryear{Kalyan, Mohta, Polozov, Batra, Jain, and
  Gulwani}{Kalyan et~al\mbox{.}}{2018}]%
        {kalyan2018neural}
\bibfield{author}{\bibinfo{person}{Ashwin Kalyan}, \bibinfo{person}{Abhishek
  Mohta}, \bibinfo{person}{Oleksandr Polozov}, \bibinfo{person}{Dhruv Batra},
  \bibinfo{person}{Prateek Jain}, {and} \bibinfo{person}{Sumit Gulwani}.}
  \bibinfo{year}{2018}\natexlab{}.
\newblock \showarticletitle{Neural-guided deductive search for real-time
  program synthesis from examples}.
\newblock \bibinfo{journal}{\emph{arXiv preprint arXiv:1804.01186}}
  (\bibinfo{year}{2018}).
\newblock


\bibitem[\protect\citeauthoryear{Kneuss, Kuraj, Kuncak, and Suter}{Kneuss
  et~al\mbox{.}}{2013}]%
        {Kneuss2013synthesis}
\bibfield{author}{\bibinfo{person}{Etienne Kneuss}, \bibinfo{person}{Ivan
  Kuraj}, \bibinfo{person}{Viktor Kuncak}, {and} \bibinfo{person}{Philippe
  Suter}.} \bibinfo{year}{2013}\natexlab{}.
\newblock \showarticletitle{Synthesis Modulo Recursive Functions}.
\newblock \bibinfo{journal}{\emph{SIGPLAN Not.}} \bibinfo{volume}{48},
  \bibinfo{number}{10} (\bibinfo{date}{Oct.} \bibinfo{year}{2013}),
  \bibinfo{pages}{407--426}.
\newblock
\showISSN{0362-1340}


\bibitem[\protect\citeauthoryear{Koukoutos, Kneuss, and Kuncak}{Koukoutos
  et~al\mbox{.}}{2016}]%
        {KoukoutosKK16}
\bibfield{author}{\bibinfo{person}{Manos Koukoutos}, \bibinfo{person}{Etienne
  Kneuss}, {and} \bibinfo{person}{Viktor Kuncak}.}
  \bibinfo{year}{2016}\natexlab{}.
\newblock \showarticletitle{An Update on Deductive Synthesis and Repair in the
  Leon Tool}. In \bibinfo{booktitle}{\emph{Proceedings Fifth Workshop on
  Synthesis, SYNT@CAV 2016, Toronto, Canada, July 17-18, 2016}}.
  \bibinfo{pages}{100--111}.
\newblock


\bibitem[\protect\citeauthoryear{Koukoutos, Raghothaman, Kneuss, and
  Kuncak}{Koukoutos et~al\mbox{.}}{2017}]%
        {koukoutos2017repair}
\bibfield{author}{\bibinfo{person}{Manos Koukoutos}, \bibinfo{person}{Mukund
  Raghothaman}, \bibinfo{person}{Etienne Kneuss}, {and} \bibinfo{person}{Viktor
  Kuncak}.} \bibinfo{year}{2017}\natexlab{}.
\newblock \showarticletitle{On repair with probabilistic attribute grammars}.
\newblock \bibinfo{journal}{\emph{arXiv preprint arXiv:1707.04148}}
  (\bibinfo{year}{2017}).
\newblock


\bibitem[\protect\citeauthoryear{Le and Gulwani}{Le and Gulwani}{2014}]%
        {PLDI-2014-LeG}
\bibfield{author}{\bibinfo{person}{Vu Le} {and} \bibinfo{person}{Sumit
  Gulwani}.} \bibinfo{year}{2014}\natexlab{}.
\newblock \showarticletitle{{FlashExtract: a framework for data extraction by
  examples}}. In \bibinfo{booktitle}{\emph{{Proceedings of the 35th Conference
  on Programming Language Design and Implementation}}},
  \bibfield{editor}{\bibinfo{person}{Michael F.~P. O'Boyle} {and}
  \bibinfo{person}{Keshav Pingali}} (Eds.). \bibinfo{publisher}{{ACM}},
  \bibinfo{pages}{55}.
\newblock
\showISBNx{978-1-4503-2784-8}
\urldef\tempurl%
\url{https://doi.org/10.1145/2594291.2594333}
\showDOI{\tempurl}


\bibitem[\protect\citeauthoryear{Lee, Heo, Alur, and Naik}{Lee
  et~al\mbox{.}}{2018}]%
        {lee2018accelerating}
\bibfield{author}{\bibinfo{person}{Woosuk Lee}, \bibinfo{person}{Kihong Heo},
  \bibinfo{person}{Rajeev Alur}, {and} \bibinfo{person}{Mayur Naik}.}
  \bibinfo{year}{2018}\natexlab{}.
\newblock \showarticletitle{Accelerating search-based program synthesis using
  learned probabilistic models}.
\newblock \bibinfo{journal}{\emph{ACM SIGPLAN Notices}} \bibinfo{volume}{53},
  \bibinfo{number}{4} (\bibinfo{year}{2018}), \bibinfo{pages}{436--449}.
\newblock


\bibitem[\protect\citeauthoryear{Menon, Tamuz, Gulwani, Lampson, and
  Kalai}{Menon et~al\mbox{.}}{2013}]%
        {menon2013machine}
\bibfield{author}{\bibinfo{person}{Aditya Menon}, \bibinfo{person}{Omer Tamuz},
  \bibinfo{person}{Sumit Gulwani}, \bibinfo{person}{Butler Lampson}, {and}
  \bibinfo{person}{Adam Kalai}.} \bibinfo{year}{2013}\natexlab{}.
\newblock \showarticletitle{A machine learning framework for programming by
  example}. In \bibinfo{booktitle}{\emph{International Conference on Machine
  Learning}}. \bibinfo{pages}{187--195}.
\newblock


\bibitem[\protect\citeauthoryear{Osera and Zdancewic}{Osera and
  Zdancewic}{2015}]%
        {osera2015type}
\bibfield{author}{\bibinfo{person}{Peter-Michael Osera} {and}
  \bibinfo{person}{Steve Zdancewic}.} \bibinfo{year}{2015}\natexlab{}.
\newblock \showarticletitle{Type-and-example-directed program synthesis}. In
  \bibinfo{booktitle}{\emph{ACM SIGPLAN Notices}}, Vol.~\bibinfo{volume}{50}.
  ACM, \bibinfo{pages}{619--630}.
\newblock


\bibitem[\protect\citeauthoryear{Peleg and Polikarpova}{Peleg and
  Polikarpova}{2020}]%
        {peleg2020perfect}
\bibfield{author}{\bibinfo{person}{Hila Peleg} {and} \bibinfo{person}{Nadia
  Polikarpova}.} \bibinfo{year}{2020}\natexlab{}.
\newblock \showarticletitle{Perfect is the Enemy of Good: Best-Effort Program
  Synthesis}. In \bibinfo{booktitle}{\emph{34th European Conference on
  Object-Oriented Programming, ECOOP}}.
\newblock


\bibitem[\protect\citeauthoryear{Perelman, Gulwani, Grossman, and
  Provost}{Perelman et~al\mbox{.}}{2014}]%
        {perelman2014test}
\bibfield{author}{\bibinfo{person}{Daniel Perelman}, \bibinfo{person}{Sumit
  Gulwani}, \bibinfo{person}{Dan Grossman}, {and} \bibinfo{person}{Peter
  Provost}.} \bibinfo{year}{2014}\natexlab{}.
\newblock \showarticletitle{Test-driven synthesis}.
\newblock \bibinfo{journal}{\emph{ACM Sigplan Notices}} \bibinfo{volume}{49},
  \bibinfo{number}{6} (\bibinfo{year}{2014}), \bibinfo{pages}{408--418}.
\newblock


\bibitem[\protect\citeauthoryear{Phothilimthana, Thakur, Bodik, and
  Dhurjati}{Phothilimthana et~al\mbox{.}}{2016}]%
        {bodik2016superoptimizaiton}
\bibfield{author}{\bibinfo{person}{Phitchaya~Mangpo Phothilimthana},
  \bibinfo{person}{Aditya Thakur}, \bibinfo{person}{Rastislav Bodik}, {and}
  \bibinfo{person}{Dinakar Dhurjati}.} \bibinfo{year}{2016}\natexlab{}.
\newblock \showarticletitle{Scaling up Superoptimization}.
\newblock \bibinfo{journal}{\emph{SIGARCH Comput. Archit. News}}
  \bibinfo{volume}{44}, \bibinfo{number}{2} (\bibinfo{date}{March}
  \bibinfo{year}{2016}), \bibinfo{pages}{297--310}.
\newblock
\showISSN{0163-5964}
\urldef\tempurl%
\url{https://doi.org/10.1145/2980024.2872387}
\showDOI{\tempurl}


\bibitem[\protect\citeauthoryear{Raychev, Vechev, and Yahav}{Raychev
  et~al\mbox{.}}{2014}]%
        {raychev2014code}
\bibfield{author}{\bibinfo{person}{Veselin Raychev}, \bibinfo{person}{Martin
  Vechev}, {and} \bibinfo{person}{Eran Yahav}.}
  \bibinfo{year}{2014}\natexlab{}.
\newblock \showarticletitle{Code completion with statistical language models}.
  In \bibinfo{booktitle}{\emph{ACM SIGPLAN Notices}},
  Vol.~\bibinfo{volume}{49}. ACM, \bibinfo{pages}{419--428}.
\newblock


\bibitem[\protect\citeauthoryear{Reynolds, Barbosa, N{\"o}tzli, Barrett, and
  Tinelli}{Reynolds et~al\mbox{.}}{2019}]%
        {reynolds2019cvc}
\bibfield{author}{\bibinfo{person}{Andrew Reynolds}, \bibinfo{person}{Haniel
  Barbosa}, \bibinfo{person}{Andres N{\"o}tzli}, \bibinfo{person}{Clark
  Barrett}, {and} \bibinfo{person}{Cesare Tinelli}.}
  \bibinfo{year}{2019}\natexlab{}.
\newblock \showarticletitle{cvc 4 sy: smart and fast term enumeration for
  syntax-guided synthesis}. In \bibinfo{booktitle}{\emph{International
  Conference on Computer Aided Verification}}. Springer,
  \bibinfo{pages}{74--83}.
\newblock


\bibitem[\protect\citeauthoryear{Shah, Kulal, and Bodik}{Shah
  et~al\mbox{.}}{2018}]%
        {shahscalable}
\bibfield{author}{\bibinfo{person}{Rohin Shah}, \bibinfo{person}{Sumith Kulal},
  {and} \bibinfo{person}{Rastislav Bodik}.} \bibinfo{year}{2018}\natexlab{}.
\newblock \showarticletitle{Scalable Synthesis with Symbolic Syntax Graphs}.
\newblock  (\bibinfo{year}{2018}).
\newblock


\bibitem[\protect\citeauthoryear{Shi, Steinhardt, and Liang}{Shi
  et~al\mbox{.}}{2019}]%
        {shi2019frangel}
\bibfield{author}{\bibinfo{person}{Kensen Shi}, \bibinfo{person}{Jacob
  Steinhardt}, {and} \bibinfo{person}{Percy Liang}.}
  \bibinfo{year}{2019}\natexlab{}.
\newblock \showarticletitle{FrAngel: component-based synthesis with control
  structures}.
\newblock \bibinfo{journal}{\emph{Proceedings of the ACM on Programming
  Languages}} \bibinfo{volume}{3}, \bibinfo{number}{POPL}
  (\bibinfo{year}{2019}), \bibinfo{pages}{1--29}.
\newblock
\urldef\tempurl%
\url{https://dl.acm.org/doi/10.1145/3290386}
\showURL{%
\tempurl}


\bibitem[\protect\citeauthoryear{Si, Yang, Dai, Naik, and Song}{Si
  et~al\mbox{.}}{2019}]%
        {si2018learning}
\bibfield{author}{\bibinfo{person}{Xujie Si}, \bibinfo{person}{Yuan Yang},
  \bibinfo{person}{Hanjun Dai}, \bibinfo{person}{Mayur Naik}, {and}
  \bibinfo{person}{Le Song}.} \bibinfo{year}{2019}\natexlab{}.
\newblock \bibinfo{title}{Learning a Meta-Solver for Syntax-Guided Program
  Synthesis}.
\newblock
\newblock
\urldef\tempurl%
\url{https://openreview.net/forum?id=Syl8Sn0cK7}
\showURL{%
\tempurl}


\bibitem[\protect\citeauthoryear{Smith and Albarghouthi}{Smith and
  Albarghouthi}{2019}]%
        {SmithA19Equivalence}
\bibfield{author}{\bibinfo{person}{Calvin Smith} {and} \bibinfo{person}{Aws
  Albarghouthi}.} \bibinfo{year}{2019}\natexlab{}.
\newblock \showarticletitle{Program Synthesis with Equivalence Reduction}. In
  \bibinfo{booktitle}{\emph{Verification, Model Checking, and Abstract
  Interpretation - 20th International Conference, {VMCAI} 2019, Cascais,
  Portugal, January 13-15, 2019, Proceedings}}. \bibinfo{pages}{24--47}.
\newblock
\urldef\tempurl%
\url{https://doi.org/10.1007/978-3-030-11245-5\_2}
\showDOI{\tempurl}


\bibitem[\protect\citeauthoryear{Solar-Lezama, Tancau, Bodik, Seshia, and
  Saraswat}{Solar-Lezama et~al\mbox{.}}{2006}]%
        {solar2006combinatorial}
\bibfield{author}{\bibinfo{person}{Armando Solar-Lezama},
  \bibinfo{person}{Liviu Tancau}, \bibinfo{person}{Rastislav Bodik},
  \bibinfo{person}{Sanjit Seshia}, {and} \bibinfo{person}{Vijay Saraswat}.}
  \bibinfo{year}{2006}\natexlab{}.
\newblock \showarticletitle{Combinatorial sketching for finite programs}.
\newblock \bibinfo{journal}{\emph{ACM SIGOPS Operating Systems Review}}
  \bibinfo{volume}{40}, \bibinfo{number}{5} (\bibinfo{year}{2006}),
  \bibinfo{pages}{404--415}.
\newblock


\bibitem[\protect\citeauthoryear{Udupa, Raghavan, Deshmukh, Mador-Haim, Martin,
  and Alur}{Udupa et~al\mbox{.}}{2013}]%
        {udupa2013transit}
\bibfield{author}{\bibinfo{person}{Abhishek Udupa}, \bibinfo{person}{Arun
  Raghavan}, \bibinfo{person}{Jyotirmoy~V Deshmukh}, \bibinfo{person}{Sela
  Mador-Haim}, \bibinfo{person}{Milo~MK Martin}, {and} \bibinfo{person}{Rajeev
  Alur}.} \bibinfo{year}{2013}\natexlab{}.
\newblock \showarticletitle{TRANSIT: specifying protocols with concolic
  snippets}.
\newblock \bibinfo{journal}{\emph{ACM SIGPLAN Notices}} \bibinfo{volume}{48},
  \bibinfo{number}{6} (\bibinfo{year}{2013}), \bibinfo{pages}{287--296}.
\newblock


\bibitem[\protect\citeauthoryear{Wang, Cheung, and Bodik}{Wang
  et~al\mbox{.}}{2017a}]%
        {wang2017synthesizing}
\bibfield{author}{\bibinfo{person}{Chenglong Wang}, \bibinfo{person}{Alvin
  Cheung}, {and} \bibinfo{person}{Rastislav Bodik}.}
  \bibinfo{year}{2017}\natexlab{a}.
\newblock \showarticletitle{Synthesizing highly expressive SQL queries from
  input-output examples}. In \bibinfo{booktitle}{\emph{Proceedings of the 38th
  ACM SIGPLAN Conference on Programming Language Design and Implementation}}.
  ACM, \bibinfo{pages}{452--466}.
\newblock


\bibitem[\protect\citeauthoryear{Wang, Dillig, and Singh}{Wang
  et~al\mbox{.}}{2017c}]%
        {Wang2017AbstractionRefinement}
\bibfield{author}{\bibinfo{person}{Xinyu Wang}, \bibinfo{person}{Isil Dillig},
  {and} \bibinfo{person}{Rishabh Singh}.} \bibinfo{year}{2017}\natexlab{c}.
\newblock \showarticletitle{Program Synthesis Using Abstraction Refinement}.
\newblock \bibinfo{journal}{\emph{Proc. ACM Program. Lang.}}
  \bibinfo{volume}{2}, \bibinfo{number}{POPL}, Article \bibinfo{articleno}{63},
  \bibinfo{numpages}{30}~pages.
\newblock
\showISSN{2475-1421}
\urldef\tempurl%
\url{https://doi.org/10.1145/3158151}
\showDOI{\tempurl}


\bibitem[\protect\citeauthoryear{Wang, Dillig, and Singh}{Wang
  et~al\mbox{.}}{2017b}]%
        {Wang2017FiniteTreeAutomata}
\bibfield{author}{\bibinfo{person}{Xinyu Wang}, \bibinfo{person}{Isil Dillig},
  {and} \bibinfo{person}{Rishabh Singh}.} \bibinfo{year}{2017}\natexlab{b}.
\newblock \showarticletitle{Synthesis of Data Completion Scripts Using Finite
  Tree Automata}.
\newblock \bibinfo{journal}{\emph{Proc. ACM Program. Lang.}}
  \bibinfo{volume}{1}, \bibinfo{number}{OOPSLA}, Article
  \bibinfo{articleno}{62}, \bibinfo{numpages}{26}~pages.
\newblock
\showISSN{2475-1421}
\urldef\tempurl%
\url{https://doi.org/10.1145/3133886}
\showDOI{\tempurl}


\bibitem[\protect\citeauthoryear{Warren}{Warren}{2013}]%
        {warren2013hacker}
\bibfield{author}{\bibinfo{person}{Henry~S Warren}.}
  \bibinfo{year}{2013}\natexlab{}.
\newblock \bibinfo{booktitle}{\emph{Hacker's delight}}.
\newblock \bibinfo{publisher}{Pearson Education}.
\newblock


\end{thebibliography}
	\clearpage
	\appendix
	\iflong
	\section*{Appendix}
	\begin{figure}[h]
	\begin{center}
		\begin{align*}
		\nonterm{Start} &\rightarrow && S &&\\
		S &\rightarrow &&\ \scode{arg0}\ |\  \scode{arg1} |\ \dots && \quad\text{string variables}\\
		&\mid&& \scode{lit-1 } |\ \scode{lit-2 } |\ \dots && \quad\text{string literals}\\
		&\mid &&\ (\scode{replace}\ S\ S\ S) && \quad\text{\scode{replace s x y} replaces first occurrence of \scode{x} in \scode{s} with \scode{y}}\\
		&\mid &&\ (\scode{concat}\ S\ S) && \quad\text{\scode{concat x y} concatenates \scode{x} and \scode{y}}\\
		&\mid &&\ (\scode{substr}\ S\ I\ I) && \quad\text{\scode{substr x y z} extracts substring of length \scode{z}, from index \scode{y}}\\
		&\mid &&\ (\scode{ite}\ B\ S\ S) && \quad\text{\scode{ite x y z} returns \scode{y} if \scode{x} is true, otherwise \scode{z}}\\
		&\mid &&\ (\scode{int.to.str}\ I) && \quad\text{\scode{int.to.str x} converts int \scode{x} to a string}\\	
		&\mid &&\ (\scode{at}\ S\ I) && \quad\text{\scode{at x y} returns the character at index \scode{y} in string {x}}\\
		B &\rightarrow &&\ \scode{true} \mid \scode{false} && \quad\text{bool literals}\\
		&\mid &&\ (\scode{=}\ I\ I) && \quad\text{\scode{= x y} returns true if \scode{x} equals \scode{y}}\\	
		&\mid &&\ (\scode{contains}\ S\ S) && \quad\text{\scode{contains x y} returns true if \scode{x} contains \scode{y}}\\	
		&\mid &&\ (\scode{suffixof}\ S\ S) && \quad\text{\scode{suffixof x y} returns true if \scode{x} is the suffix of \scode{y}}\\
		&\mid &&\ (\scode{prefixof}\ S\ S) && \quad\text{\scode{prefixof x y} returns true if \scode{x} is the prefix of \scode{y}}\\
		I &\rightarrow &&\ \scode{arg0}\ |\  \scode{arg1} |\ \dots && \quad\text{int variables}\\
		&\mid&& \scode{lit-1 } |\ \scode{lit-2 } |\ \dots && \quad\text{int literals}\\
		&\mid &&\ (\scode{str.to.int}\ S) && \quad\text{\scode{str.to.int x} converts string x to a int}\\	
		&\mid &&\ (\scode{+}\ I\ I) && \quad\text{\scode{+ x y} sums \scode{x} and \scode{y}}\\
		&\mid &&\ (\scode{-}\ I\ I) && \quad\text{\scode{- x y} subtracts \scode{y} from \scode{x}}\\
		&\mid &&\ (\scode{length}\ S) && \quad\text{\scode{length x} returns length of \scode{x}}\\
		&\mid &&\ (\scode{ite}\ B\ I\ I) && \quad\text{\scode{ite x y z} returns \scode{y} if \scode{x} is true, otherwise \scode{z}}\\
		&\mid &&\ (\scode{indexof}\ S\ S\ I) && \quad\text{\scode{indexof x y z} returns index of \scode{y} in \scode{x}, starting at index \scode{z}}\\
		\end{align*}
	\end{center}
	\caption{The full \sygus \stringbench grammar of the \euphony benchmark suite. Integer and string variables and constants change per benchmark. Some benchmark files contain a reduced grammar.}\label{fig:eval-grammar}
\end{figure}%

\begin{figure}[h]
	\begin{center}
		\begin{align*}
		\nonterm{Start} &\rightarrow && BV &&\\
		BV &\rightarrow &&\ \scode{arg0}\ |\  \scode{arg1} |\ \dots && \quad\text{bit-vector variables}\\
		&\mid&& \scode{lit-1 } |\ \scode{lit-2 } |\ \dots && \quad\text{bit-vector literals}\\
		&\mid &&\ (\scode{xor}\ BV\ BV) && \quad\text{\scode{xor x y} performs bitwise xor between \scode{x} and \scode{y}}\\
		&\mid &&\ (\scode{and}\ BV\ BV) && \quad\text{\scode{and x y} performs bitwise and operation between \scode{x} and \scode{y}}\\
		&\mid &&\ (\scode{or}\ BV\ BV) && \quad\text{\scode{or x y} performs bitwise or operation between \scode{x} and \scode{y}}\\
		&\mid &&\ (\scode{neg}\ BV) && \quad\text{\scode{neg x} returns the two's complement of \scode{x}}\\
		&\mid &&\ (\scode{not}\ BV) && \quad\text{\scode{not x} returns the one's complement of \scode{x}}\\
		&\mid &&\ (\scode{add}\ BV\ BV) && \quad\text{\scode{add \scode{x} \scode{y}} adds \scode{x} and \scode{y}}\\	
		&\mid &&\ (\scode{mul}\ BV\ BV) && \quad\text{\scode{mul \scode{x} \scode{y}} multiplies \scode{x} and \scode{y}}\\
		&\mid &&\ (\scode{udiv}\ BV\ BV) && \quad\text{\scode{udiv \scode{x} \scode{y}} returns the unsigned quotient of dividing \scode{x} by \scode{y}}\\
		&\mid &&\ (\scode{urem}\ BV\ BV) && \quad\text{\scode{urem x y} returns the unsigned remainder of dividing \scode{x} by \scode{y}}\\
		&\mid &&\ (\scode{lshr}\ BV\ BV) && \quad\text{\scode{lshr x y} returns the logical right shift of \scode{x} by \scode{y} bits}\\
		&\mid &&\ (\scode{ashr}\ BV\ BV) && \quad\text{\scode{ashr x y} returns the arithmetic right shift of \scode{x} by \scode{y}}\\
		&\mid &&\ (\scode{shl}\ BV\ BV) && \quad\text{\scode{shl x y} returns the logical left shift of \scode{x} by \scode{y}}\\
		&\mid &&\ (\scode{sdiv}\ BV\ BV) && \quad\text{\scode{sdiv \scode{x} {y}} returns the signed quotient of dividing \scode{x} by \scode{y}}\\
		&\mid &&\ (\scode{srem}\ BV\ BV) && \quad\text{\scode{srem \scode{x} \scode{y}} returns the signed remainder of dividing \scode{x} by \scode{y}}\\
		&\mid &&\ (\scode{sub}\ BV\ BV) && \quad\text{\scode{sub \scode{x} \scode{y}} subtracts \scode{y} from \scode{x}}\\
		&\mid &&\ (\scode{ite}\ B\ BV\ BV) && \quad\text{\scode{ite x y z} returns \scode{y} if \scode{x} is true, otherwise \scode{z}}\\
		B &\rightarrow &&\ \scode{true} \mid \scode{false} && \quad\text{bool literals}\\
		&\mid &&\ (\scode{=}\ BV\ BV) && \quad\text{\scode{= x y} returns true if \scode{x} equals \scode{y}}\\	
		&\mid &&\ (\scode{ult}\ BV\ BV) && \quad\text{\scode{ult x y} returns true if  \scode{x} is unsigned less than \scode{y}}\\	
		&\mid &&\ (\scode{ule}\ BV\ BV) && \quad\text{\scode{ule x y} returns true if \scode{x} is unsigned less than equal to \scode{y}}\\	
		&\mid &&\ (\scode{slt}\ BV\ BV) && \quad\text{\scode{slt x y} returns true if \scode{x} is signed less than \scode{y}}\\
		&\mid &&\ (\scode{sle}\ BV\ BV) && \quad\text{\scode{sle x y} returns true if \scode{x} is signed less than equal to \scode{y}}\\
		&\mid &&\ (\scode{ugt}\ BV\ BV) && \quad\text{\scode{ugt x y} returns true if \scode{x} unsigned greater than \scode{y}}\\
		&\mid &&\ (\scode{redor}\ BV) && \quad\text{\scode{redor x} performs bit-wise or reduction of \scode{x}}\\
		&\mid &&\ (\scode{and}\ BV\ BV) && \quad\text{\scode{and x y} returns the logical and of \scode{x} and \scode{y}}\\
		&\mid &&\ (\scode{or}\ BV\ BV) && \quad\text{\scode{or x y} returns the logical or of \scode{x} and \scode{y}}\\
		&\mid &&\ (\scode{not}\ BV) && \quad\text{\scode{not x} returns the logical not of \scode{x}}\\
		\end{align*}
	\end{center}
	\caption{The full \sygus \bitvec grammar of the Hacker's Delight benchmarks; variables and constants change per benchmark. Some of the benchmarks contain a reduced grammar; required constants are provided.}\label{fig:eval-grammar-bitvec}
\end{figure}%

\begin{figure}[h]
	\begin{center}
		\begin{align*}
		\nonterm{Start} &\rightarrow && B &&\\	
		B &\rightarrow &&\ 	\scode{arg0 } |\ \scode{arg1 } |\ \dots && \quad\text{boolean variables}\\
		&\mid &&\ (\scode{and}\ B\ B) && \quad\text{\scode{and x y} returns the logical and of \scode{x} and \scode{y}}\\
		&\mid &&\ (\scode{not}\ B) && \quad\text{\scode{not x} returns the logical not of \scode{x}}\\
		&\mid &&\ (\scode{or}\ B\ B) && \quad\text{\scode{or x y} returns the logical or of \scode{x} and \scode{y}}\\
		&\mid &&\ (\scode{xor}\ B\ B) && \quad\text{\scode{xor x y} returns the logical xor of \scode{x} and \scode{y}}\\	
		\end{align*}
	\end{center}
	\caption{The full \sygus \circuit grammar of the \euphony benchmark suite. Variables and the depth of the grammar change per benchmark.}\label{fig:eval-grammar-circuit}
\end{figure}%

	\fi

\end{document}